# A Mechanism-Coupled Split Window Network for Medium- to High-Resolution Land Surface Temperature Retrieval

***Tian Xie*[a], *Menghui Jiang*[a], *Chao Zeng*[a], *Huifang Li*[a], *Guanhao Zhang*[a], *Chan Li*[a], *Huanfeng Shen*[a,b,c*]**

*[a] School of Resource and Environmental Sciences, Wuhan University, Wuhan 430079, China*

*[b] Key Laboratory of Geographic Information System of Ministry of Education, Wuhan 430079, China*

*[c] Key Laboratory of Digital Cartography and Land Information Application of the Ministry of Natural Resources, Wuhan 430079, China*

***Corresponding author.**

*E-mail address:* shenhf@whu.edu.cn (H. Shen)

**Abstract**

Land surface temperature (LST) is a fundamental physical variable in land-atmosphere interactions, surface energy budgets, and climate processes. LST derived from medium- to high-resolution thermal infrared (TIR) observations effectively reveals thermal environmental disparities across distinct landscape units. However, achieving accurate, robust, and globally generalizable LST retrieval remains challenging under complex atmospheric conditions and diverse land cover types. Traditional split window (SW) algorithms heavily rely on empirical parameterizations, whose fixed coefficients fail to adapt to complex scenarios such as high surface temperatures and high atmospheric water vapor content. Concurrently, conventional data-driven models exhibit limited generalizability to out-of-distribution (OOD) samples due to the absence of explicit physical structure constraints. To address these issues, this study proposes a Parallel Component Decoupled Neural Network (PCD-Net) framework, which reformulates SW retrieval as a dynamic learning problem of physical component coefficients. Using the SW equation as the physical backbone, the framework constructs parallel subnetworks to adaptively learn the dynamic coefficients corresponding to the constant, first-order, and second-order brightness temperature difference terms; meanwhile, a residual branch is incorporated to supplement the nonlinear coupling corrections induced by the joint effects of surface emissivity and atmospheric water vapor. Through this component-level decoupled modeling, PCD-Net explicitly characterizes the dynamic response relationships between land surface emissivity, atmospheric water vapor content, and different SW physical components. PCD-Net is first initialized and jointly optimized using radiative transfer simulation samples, and is subsequently fine-tuned using global in situ LST observations, enabling the model to effectively adapt from simulated environments to real observational conditions. Validation using 4,450 in situ samples from 29 globally distributed benchmark sites shows that PCD-Net achieves a mean absolute error (MAE) of 1.84 K, a root mean square error (RMSE) of 2.55 K, and a coefficient of determination ($R^2$) of 0.966, outperforming the traditional split window mechanistic model (MM(SW)), the radiative transfer mechanistic model (MM(RT)), and the ML baseline. Under extreme conditions, including extremely low/high water vapor and extremely low/high LST, PCD-Net substantially reduces retrieval errors relative to the traditional MM(SW), with error reductions exceeding 50% in some extreme scenarios, demonstrating stronger cross-scenario robustness. Sensitivity analysis further indicates that PCD-Net exhibits smoother responses to perturbations in at-sensor radiance, atmospheric water vapor content, and land surface emissivity, suggesting lower sensitivity to input uncertainties.

**Keywords:** Land surface temperature; Split window algorithm; Mechanistic model; Machine learning; Component decoupling.

## 1. Introduction

Land surface temperature (LST) is a key parameter governing land–atmosphere interactions, surface energy budgets, and climate change processes (Li et al., 2013; Li et al., 2023). It plays a fundamental role in climate change monitoring (Li et al., 2025; Li and Thompson, 2021), urban heat island assessment (Shen et al., 2016), water cycle modeling (Anderson et al., 2012), and extreme heatwave warning (Gu et al., 2025). LST has not only been explicitly identified as a priority core parameter by the International

Geosphere-Biosphere Programme (IGBP) (Townshend et al., 1994), but has also been incorporated into the Earth System Data Record (ESDR) program of the National Aeronautics and Space Administration (NASA) (Hollmann et al., 2013).

However, existing satellite thermal infrared (TIR) observation systems face a persistent trade-off between spatial resolution and spatiotemporal continuity (Wu et al., 2021; Weng et al., 2014; Zhan et al., 2013). Coarse-resolution sensors such as MODIS and VIIRS can provide long term and stable global LST products (Hulley et al., 2018; Wan, 2014, 2008), but their kilometer scale pixels are insufficient to characterize substantial thermal variations within urban built up areas, agricultural mosaics, and natural vegetation patches (Weng, 2009; Li et al., 2023; Zhan et al., 2013; Bian et al., 2024). In contrast, very high spatial resolution TIR observations are constrained by limited spatial coverage and revisit frequency, making them difficult to support long term continuous analyses at the global scale (Wu et al., 2021; Weng et al., 2014). In this context, medium- to high-spatial-resolution TIR data at 30–100 m, represented by Landsat, ASTER, and ECOSTRESS, have become an important scale for linking local surface processes with global change studies, owing to their strong capability in capturing landscape scale thermal differentiation and their potential for wide area coverage (Wu et al., 2021; Weng et al., 2014). Nevertheless, under complex global atmospheric backgrounds and diverse land cover conditions, achieving accurate and robust medium- to high-resolution LST retrieval while maintaining physical plausibility remains a key challenge in TIR remote sensing (Li et al., 2023; Jimenez-Munoz et al., 2014; Hulley et al., 2012; Cao et al., 2021).

In satellite TIR remote sensing, atmospheric absorption and re-emission make LST retrieval an inherently complex and ill-posed inverse problem (Li et al., 2013; Li et al., 2023). To address this challenge, various mechanism-driven LST retrieval algorithms have been developed by introducing physical approximations or simplifications to the radiative transfer equation (RTE) (Gillespie et al., 2011; Jimenez-Munoz et al., 2014; Jimenez-Munoz and Sobrino, 2008; Qin et al., 2001; Jimenez-Munoz and Sobrino, 2003; Jimenez-Munoz et al., 2009; Rozenstein et al., 2014; Yu et al., 2014b; Chen et al., 2015; Wang et al., 2016; He et al., 2026). According to the configuration of TIR observation channels, existing methods can generally be classified into single channel, dual channel, and multichannel approaches. Among these, the dual channel split window (SW) algorithm exploits the differential atmospheric absorption between two adjacent TIR bands to reduce the influence of atmospheric water vapor (Jimenez-Munoz et al., 2014; Jimenez-Munoz and Sobrino, 2008), which is widely adopted due to its mathematical simplicity and robust performance. However, traditional SW methods typically rely on fixed empirical coefficients or predefined empirical parameterizations, which may lead to substantial systematic biases under extreme surface and atmospheric conditions, such as high-temperature and high-humidity environments (Li et al., 2013; Yu et al., 2014b; Duan et al., 2018; Li et al., 2023). Therefore, they remain insufficient for meeting the high accuracy requirements of medium- to high-resolution LST retrieval across globally complex scenarios.

In recent years, machine learning (ML) techniques have demonstrated remarkable potential in geophysical parameter retrieval because of their strong capability for modeling complex nonlinear relationships (Yuan et al., 2020; Reichstein et al., 2019; Bergen et al., 2019). Compared with traditional empirical parameterizations, ML can automatically learn high-dimensional mapping relationships between input features and target variables from large datasets. However, data-driven models generally lack explicit physical constraints and interpretable mechanisms, and their performance may deteriorate when the training samples do not adequately represent specific climate zones, land cover types, or extreme conditions (Koldasbayeva et al., 2024; Reichstein et al., 2019). In such out-of-distribution (OOD) scenarios, purely data-driven retrieval models are more likely to produce physically inconsistent or unstable predictions (Shen and Zhang, 2023; Karpatne et al., 2017).

To alleviate this issue, previous studies have attempted to integrate radiative transfer mechanisms with ML through knowledge guided strategies, physical constraints, or model embedding to improve retrieval accuracy (Cheng et al., 2025; Zhang et al., 2025; Wang et al., 2021). However, existing studies still have two main limitations. First, most mechanism–learning coupled methods have been designed and validated primarily for coarse-resolution products such as MODIS and AVHRR (Guan et al., 2026; Cheng et al., 2025), with limited systematic evaluation in medium- to high-resolution scenarios. The latter are more sensitive to local surface thermal heterogeneity, land cover differences, and atmospheric variability. Second, in terms of network design, existing methods generally learn multiple split window related parameters or correction relationships through an integrated network pathway (Cheng et al., 2025; Zhang et al., 2025). This holistic learning strategy may cause nonlinear entanglement among the contributions of different physical terms within a shared feature space, thereby weakening the model's representation of specific physical processes and limiting its robustness and generalizability in globally complex scenarios.

To address these limitations, this study proposes a Parallel Component Decoupled Neural Network framework (PCD-Net) for global medium- to high-resolution LST retrieval. Using the dual-channel SW equation as the physical backbone, PCD-Net explicitly decomposes LST retrieval into a constant term, a first-order brightness temperature difference term, a second-order brightness temperature difference term, and a coupling residual term. Multiple parallel subnetworks are constructed to estimate the dynamic coefficients of these physical components, thereby transforming conventional fixed coefficient SW retrieval into an adaptive component-wise coefficient learning problem under complex surface and atmospheric conditions. In terms of training strategy, PCD-Net first uses radiative transfer simulation samples to initialize the parallel subnetworks with physical priors and further performs simulation-driven end-to-end joint optimization within the complete SW physical backbone. It is then fine-tuned in a supervised manner using global in situ LST samples to reduce discrepancies between the simulation environment and real remote sensing observations. Compared with the holistic parameter learning strategy based on a single network, PCD-Net aims to reduce the entanglement among different physical terms in a unified feature space by decoupling their learning pathways, thereby enhancing model adaptability under complex surface and atmospheric conditions.

Using Landsat 8 TIR observations as a representative medium- to high-resolution dual-channel TIR data source, this study conducts a systematic evaluation across five continents by integrating MODTRAN-driven simulation samples based on the GAPRI and TIGR atmospheric profile databases, reanalysis data, and observations from 29 ground stations worldwide. The evaluation covers diverse surface and atmospheric conditions, ranging from arid to humid environments and from tropical to high latitude regions, as well as multiple representative land cover types, including cropland, grassland, desert, wetland, tundra, and alpine meadow. In addition, extreme condition experiments and parameter sensitivity analyses are conducted to quantify model stability under boundary conditions and characterize its error response to input uncertainties, providing a methodological basis for future extension to ASTER, ECOSTRESS, and other medium- to high-resolution TIR platforms.

The main contributions of this study are threefold:

(1) A parallel component decoupled framework is proposed for dual-channel SW retrieval, reformulating traditional static empirical coefficient estimation as a dynamic coefficient learning task based on physical component decoupling, thereby enabling deep collaboration between the physical mechanism backbone and data-driven mapping.

(2) A training strategy that combines simulation-driven physical initialization, end-to-end joint optimization, and supervised fine-tuning with in situ observations is established. This strategy preserves the physical structural prior of the SW formulation while improving model adaptability to real remote sensing observations.

(3) Through global site-based validation, extreme condition testing, and input perturbation analysis, the proposed framework is systematically demonstrated to achieve stronger generalization capability and more stable performance under complex atmospheric and surface environments, providing key algorithmic support for developing globally high accuracy medium- to high-resolution LST products.

## 2. Data

### 2.1 Simulation data

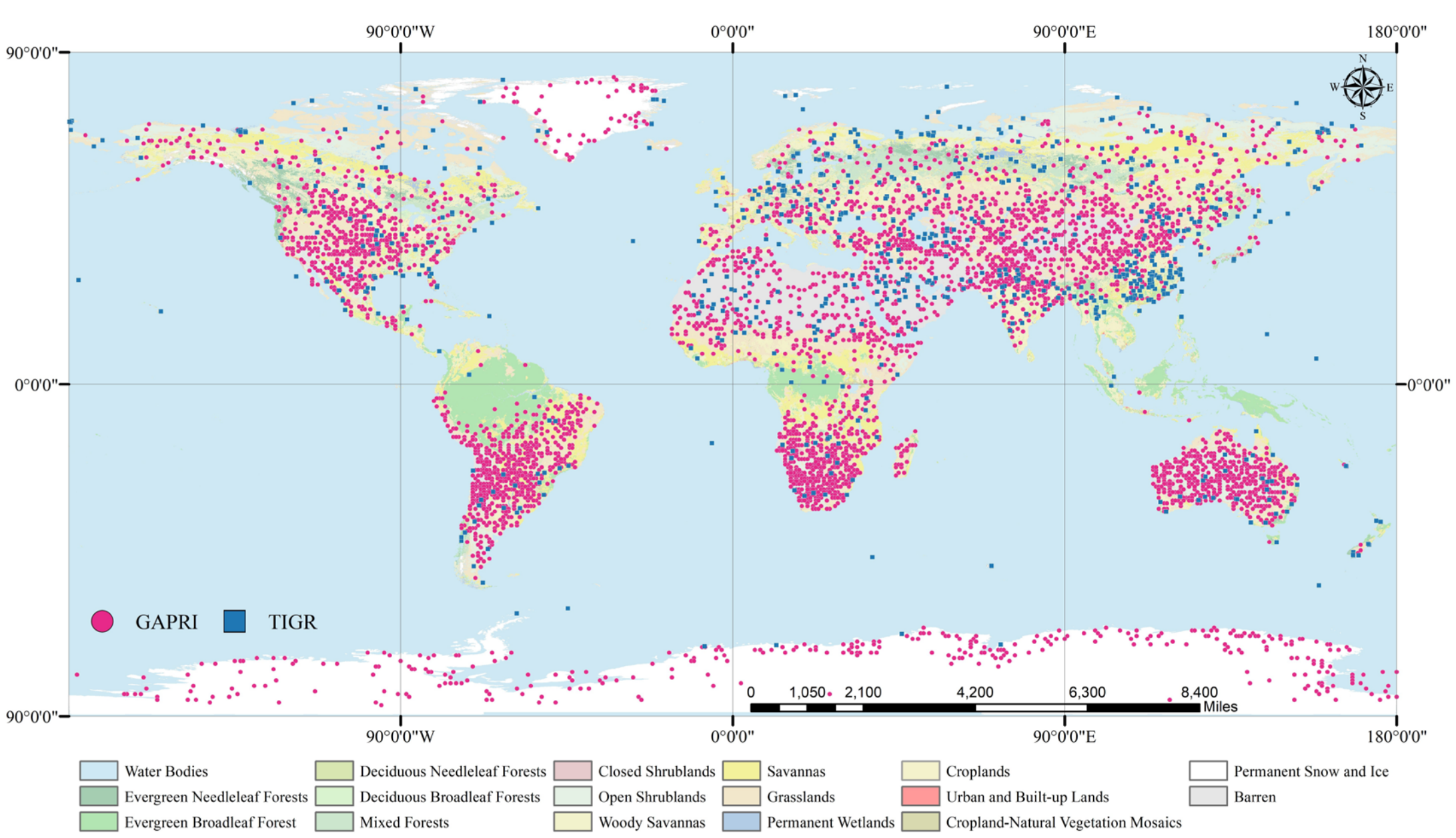

Fig. 1. Spatial distribution of the atmospheric profile datasets. Global land-area coverage of the TIGR and GAPRI atmospheric profile datasets used in this study.

To construct a globally representative simulation sample library, this study generated large-scale samples based on radiative transfer simulations. The simulation samples were mainly derived from two global atmospheric vertical profile databases: the Global Atmospheric Profile Database (GAPRI) and the Thermodynamic Initial Guess Retrieval (TIGR) database. GAPRI is a widely used atmospheric profile dataset (Mattar et al., 2015), covering the globe on an approximately 0.75° × 0.75° latitude–longitude grid with a 6 h temporal resolution. It contains 8,324 atmospheric profiles with 29 vertical levels. The dataset covers five typical atmospheric regimes from polar to tropical regions and spans a broad range of precipitable water vapor from 0 to 6.0 $g \cdot cm^{-2}$, thereby providing sufficient samples for simulations under different humidity conditions. The TIGR atmospheric profile database was constructed from more than 80,000 radiosonde observations, from which 2311 representative profiles were obtained through statistical sampling (Chedin et al., 1985; Chevallier et al., 1998). Each profile records the vertical distributions of temperature, water vapor, and ozone concentration, and is classified into five air-mass types, including tropical, mid latitude, and subarctic conditions. To avoid interference from ocean pixels in land surface modeling, only atmospheric profiles located over land areas were retained in this study. The global spatial distributions of the two atmospheric profile datasets over land areas are shown in Fig. 1.

The atmospheric radiative transfer model MODTRAN 5.2.2 (Berk et al., 2004) was used to simulate the thermal infrared radiative transfer process. MODTRAN can characterize key radiative transfer processes in the thermal infrared region, including molecular absorption, water vapor absorption, path radiance, and surface emission. The simulation outputs include thermal infrared radiance and associated atmospheric parameters, which were further used to construct the sample dataset for parallel subnetwork initialization and performance evaluation under simulated conditions.

### 2.2 Ground-based observations

This study selected three globally representative surface radiation observation networks as validation data sources: the Baseline Surface Radiation Network (BSRN), the Surface Radiation Budget Network (SURFRAD), and the Heihe Watershed Allied Telemetry Experimental Research network (HIWATER). BSRN, established by the World Radiation Monitoring Center, has deployed stations across different climate zones worldwide since 1992. It provides high accuracy and high-temporal-resolution shortwave and longwave radiation flux observations and has been widely used as a benchmark dataset for remote sensing validation and radiation model evaluation (Driemel et al., 2018). SURFRAD, operated by the National Oceanic and Atmospheric Administration (NOAA), has maintained seven stations across different climate regions of the United States since 1993. It continuously provides broadband solar radiation, thermal infrared radiation, and related meteorological observations, all of which are subject to strict quality control and released as daily aggregated datasets (Augustine et al., 2000) . The HIWATER network is located in the arid region of northwestern China and covers complex underlying surfaces such as cropland, Gobi desert, wetland, and alpine grassland (Yu et al., 2011; Yu et al., 2014a; Yu et al., 2017), providing long term and continuous ground observations for regional land surface processes and land–atmosphere water and heat flux studies. In total, 29 validation sites distributed across Asia, Europe, North America, Africa, and South America were selected in this study (Table 1). These sites cover a wide range of representative landscape types, including cropland, grassland, desert, wetland, urban surfaces, and alpine grassland. LST was estimated based on the surface longwave radiation balance using upward and downward longwave radiation fluxes measured by ground-based four-component radiometers, expressed as follows:

$$LST = \left[\frac{I\uparrow - (1-\varepsilon_b)\cdot I\downarrow}{\varepsilon_b \cdot \sigma}\right]^{1/4} \quad (1)$$

where $I\uparrow$ and $I\downarrow$ are the upward and downward longwave radiation fluxes measured at the ground surface, $\sigma$ is the Stefan-Boltzmann constant ($\sigma$ = 5.6704×10$^{-8}$ W·m$^{-2}$·K$^{-4}$), and $\varepsilon_b$ is the broadband land surface emissivity. The broadband emissivity was converted from the narrowband emissivity data of the MODIS 8-day composite LST and emissivity product (MOD11A2) (Wang et al., 2005). The MODIS image temporally closest to the Landsat overpass was selected for extraction.

**Table 1**

Summary of the global ground validation sites used in this study.

| Continent | Code | Name | Surface Type | Location | Elevation | $N_{obs}$ |
|---|---|---|---|---|---|---|
| Asia | AR | A'rou | Subalpine Meadow | 38.0473N, 100.4643E | 3033 m | 147 |
| | CAP | Cape Baranova | Tundra | 79.2700N, 101.7500E | 20–30 m | 14 |
| | DM | Daman | Maize Cropland | 38.8555N, 100.3722E | 1556 m | 168 |
| | DSL | Dashalong | Marsh Alpine Meadow | 38.8399N, 98.9406E | 3739 m | 117 |
| | HH | Heihe | Grassland | 38.827N, 100.4756E | 1560 m | 97 |
| | HMZ | Desert | *Reaumuria* Desert | 42.1135N, 100.9872E | 1054 m | 269 |
| | HZZ | Huazhaizi Desert Steppe | *Kalidium Foliatum* Desert | 38.7659N, 100.3201E | 1731 m | 167 |
| | JYL | Jingyangling | Alpine Meadow | 37.8384N, 101.1160E | 3750 m | 123 |
| | SDQ | Sidaoqiao | *Tamarix* | 42.0012N, 101.1374E | 873 m | 286 |
| | TAT | Tateno | Grassland | 36.0581N, 140.1258E | 25 m | 104 |
| | YK | Yakou | Alpine Meadow | 38.0142N, 100.2421E | 4148 m | 45 |
| | ZYSD | Zhangye wetland | Reed Wetland | 38.9751N, 100.4464E | 1460 m | 166 |
| Europe | BUD | Budapest-Lorinc | Grassland | 47.4291N, 19.1822E | 139 m | 73 |
| | CAB | Cabauw | Grassland | 51.9680N, 4.9280E | 0 m | 132 |
| | NYA | Ny-Ålesund | Tundra | 78.9227N, 11.9273E | 11 m | 163 |
| | PAY | Payerne | Cultivated Cropland | 46.8123N, 6.9422E | 491 m | 291 |
| | TOR | Toravere | Grassland | 58.2641N, 26.4613E | 70 m | 73 |
| North America | BAR | Barrow | Tundra | 71.3230N, 156.6070W | 8 m | 41 |
| | BND | Bondville | Grassland | 40.0519N, 88.3731W | 230 m | 205 |
| | DRA | Desert Rock | Desert | 36.6237N, 116.0195W | 1007 m | 337 |
| | FPK | Fort Peck | Grassland | 48.3078N, 105.1017W | 634 m | 308 |
| | GWN | Goodwin Creek | Grassland | 34.2547N, 89.8729W | 98 m | 245 |
| | PSU | Penn. State Univ | Cultivated Cropland | 40.7201N, 77.9308W | 376 m | 68 |

|  | SEL | Selegua | Grassland | 15.7840N, 91.9902W | 602 m | 29 |
|---|---|---|---|---|---|---|
|  | SXF | Sioux Falls | Grassland | 43.7340N, 96.6233W | 473 m | 214 |
|  | TBL | Table Mountain | Grassland | 40.1250N, 105.2368W | 1689 m | 258 |
| Africa | GOB | Gobabeb | Desert Gravel | 23.5614S, 15.0420E | 407 m | 216 |
|  | IZA | Izaña | Rock | 28.3093N, 16.4993W | 2373 m | 83 |
| South America | OHY | Observatory of Huancayo | Grassland | 12.0500S, 75.3200W | 3314 m | 11 |
|  |  |  | **Total** |  |  | 4,450 |

*$N_{obs}$ denotes the number of clear-sky Landsat–in situ matchup samples.

### 2.3 Satellite observations and auxiliary variables

This study used thermal infrared observations from the Landsat 8 Thermal Infrared Sensor (TIRS) as the primary satellite data source. Two TIR bands, namely TIRS Band 10 and Band 11, were selected as the core input variables for LST retrieval. Considering the stray-light effect in Landsat 8 TIRS Band 11, this study used Landsat Collection 2 data that had been recalibrated and stray-light corrected by the United States Geological Survey (USGS), thereby reducing the potential systematic errors caused by B11 stray light. The images used in this study covered Landsat 8 observations from 2013 to 2023, providing data support for systematically evaluating model applicability and stability.

In addition to TIR brightness temperatures, LST retrieval also requires auxiliary variables describing land surface and atmospheric states. Land surface emissivity was dynamically estimated by integrating the ASTER GED v3 global emissivity product with the normalized difference vegetation index (NDVI) threshold method derived from concurrent Landsat 8 reflective bands. This strategy was used to account for the influence of vegetation phenological variations on pixel-level emissivity. Atmospheric water vapor content was extracted from the National Centers for Environmental Prediction (NCEP) 6-hourly reanalysis data and linearly interpolated to the Landsat 8 overpass time, ensuring spatiotemporal consistency between auxiliary parameters and satellite observations.

## 3. Methodology

### 3.1 Physical basis

According to atmospheric radiative transfer theory, under clear-sky conditions, the at-sensor radiance received by a satellite sensor is jointly determined by surface thermal emission, atmospheric upwelling radiance, and atmospheric downwelling radiance reflected by the surface. It can be expressed as:

$$L_{sen} = \varepsilon B(T_s)\tau + (1-\varepsilon)I\downarrow\tau + I\uparrow \quad (2)$$

where $\varepsilon$ is land surface emissivity, $B(T_s)$ is the blackbody radiance corresponding to land surface temperature $T_s$. $\tau$ is the atmospheric transmittance. Retrieval methods based on the RTE can explicitly describe the radiative coupling between the land surface and the atmosphere in thermal infrared observations, and therefore provide the fundamental theoretical basis for LST retrieval. This method has a clear physical foundation and can achieve high retrieval accuracy, but it strongly depends on real-time atmospheric parameters.

To reduce the dependence on real-time atmospheric profile data, traditional split window mechanism models MM(SW) have been widely used for thermal infrared LST retrieval. These methods exploit the differential atmospheric absorption responses of two adjacent thermal infrared window bands and construct brightness temperature combinations to reduce atmospheric effects, thereby approximating the complex radiative transfer process in a parameterized form. In this study, the generalized split window algorithm proposed by Jimenez-Munoz et al. (2014) is adopted, which can be written as:

$$T_{\mathrm{s}} = T_{\mathrm{i}} + c_1(T_{\mathrm{i}} - T_{\mathrm{j}}) + c_2(T_{\mathrm{i}} - T_{\mathrm{j}})^2 + c_0 + (c_3 + c_4\omega)(1 - \bar{\varepsilon}) + (c_5 + c_6\omega)\Delta\varepsilon \tag{3}$$

where $T_i$ and $T_j$ are the brightness temperatures of two thermal infrared bands, $\bar{\varepsilon}$ is the mean emissivity, $\Delta\varepsilon$ is the emissivity difference between the two bands, $\omega$ is atmospheric water vapor content, and $c_0$ to $c_6$ are empirical coefficients fitted from simulation data. This algorithm approximates the complex radiative transfer process using a concise parameterized formulation and is computationally efficient. However, because it relies on empirical assumptions, its retrieval errors can become substantial under extreme atmospheric and surface conditions, such as high-temperature and high-humidity environments.

## 3.2 Proposed PCD-Net framework

### 3.2.1 Overall architecture

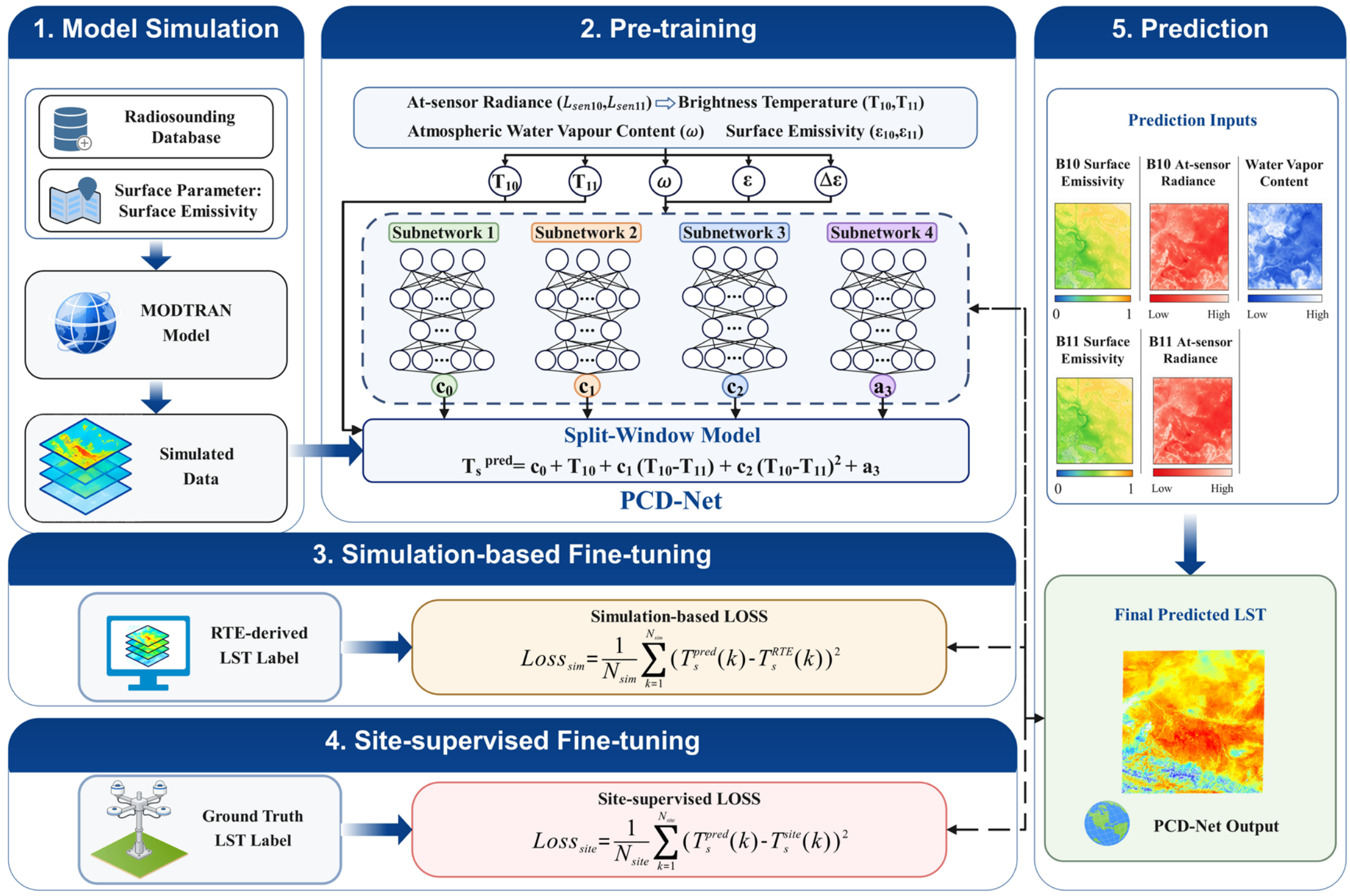

Fig. 2. Schematic illustration of the proposed Parallel Component Decoupled Neural Network (PCD-Net) framework. The framework consists of five sequential stages: radiative transfer simulation sample construction, parallel

component subnetwork pretraining, simulation-based end-to-end fine-tuning, site-supervised fine-tuning, and final LST prediction.

To overcome the limitations of fixed-coefficient SW retrieval under complex surface and atmospheric conditions, this study proposes a PCD-Net framework. Using the generalized SW algorithm as the physical backbone, PCD-Net decomposes the LST retrieval process into a constant term, a first-order brightness temperature difference term, a second-order brightness temperature difference term, and a surface–atmosphere coupling residual term. Multiple parallel subnetworks are then constructed to estimate the dynamic coefficients corresponding to these physical components. As shown in Fig. 2, the overall workflow of PCD-Net includes five stages. First, a physical simulation sample set is constructed using globally representative atmospheric profiles, land surface emissivity parameters, and MODTRAN radiative transfer simulations. Second, the parallel component subnetworks are pretrained using the simulation samples, enabling each branch to obtain an initial parameter representation corresponding to a specific SW physical component. Third, the pretrained branches are embedded into the complete SW physical backbone and further fine-tuned end-to-end using the reference LST derived from radiative transfer simulations as labels. Fourth, global in situ LST samples are used for site-supervised fine-tuning to correct discrepancies between the simulation environment and real remote sensing observations. Finally, the trained PCD-Net is applied to actual remote sensing images for LST retrieval.

Based on the atmospheric profile data and MODTRAN simulation results described in Section 2.1, this study constructs a radiative transfer sample set covering various climate and atmospheric conditions. Each sample contains land surface emissivities of Landsat 8 TIRS Band 10 and Band 11, denoted as $\varepsilon_{10}$ and $\varepsilon_{11}$, atmospheric water vapor content $\omega$, and the corresponding at-sensor radiances $L_{10}$ and $L_{11}$. The brightness temperatures $T_{10}$ and $T_{11}$ are calculated using the inverse Planck function:

$$T_i = \frac{K_{2,i}}{\ln\left(\frac{K_{1,i}}{L_i} + 1\right)}, \quad i = 10,11 \tag{4}$$

where $K_{1,i}$ and $K_{2,i}$ are the radiometric calibration constants of the i-th Landsat 8 TIRS thermal infrared band. Atmospheric water vapor content $\omega$ is obtained by vertically integrating the water vapor density in the atmospheric profile:

$$\omega = \int_0^{\infty} \rho_v(z)dz \tag{5}$$

where $\rho_v(z)$ is the water vapor density at altitude $z$. The mean emissivity $\bar{\varepsilon}$ and emissivity difference $\Delta\varepsilon$ are further defined as:

$$\bar{\varepsilon} = \frac{\varepsilon_{10} + \varepsilon_{11}}{2} \tag{6}$$

$$\Delta\varepsilon = \varepsilon_{10} - \varepsilon_{11} \tag{7}$$

Accordingly, the input variables of PCD-Net can be expressed as $T_{10}$, $T_{11}$, $\omega$, $\bar{\varepsilon}$, and $\Delta\varepsilon$. Among them, $T_{10}$ and $T_{11}$ are retained in the SW physical backbone to represent the dual-channel thermal infrared brightness temperature response, whereas $\omega$, $\bar{\varepsilon}$, and $\Delta\varepsilon$ are used as inputs to the parallel subnetworks for dynamically estimating the parameters of different physical components.

#### 3.2.2 Parallel learning of SW parameters

In the generalized SW formulation, the brightness temperatures and their difference terms participate in LST retrieval through a fixed analytical structure, while the coefficients $c_0$–$c_2$ are mainly modulated by atmospheric water vapor and land surface emissivity conditions. Based on this characteristic, PCD-Net retains the brightness temperature difference structure as a fixed physical backbone. Instead of directly performing holistic black-box regression from inputs to LST, it dynamically learns key parameters in the SW equation through parallel subnetworks. Specifically, the first three subnetworks take $\omega$, $\bar{\varepsilon}$ and $\Delta\varepsilon$ as inputs and predict the coefficients corresponding to the constant term, the first-order brightness temperature difference term, and the second-order brightness temperature difference term in the SW equation:

$$c_i = f_{subnet,i}(\omega, \bar{\varepsilon}, \Delta\varepsilon), \;\; i = 0, 1, 2 \tag{8}$$

where $c_0$, $c_1$, and $c_2$ correspond to the regression coefficients of the constant term, first-order brightness temperature difference term, and second-order brightness temperature difference term in the classical SW formulation, respectively. Through this parameterized mapping, PCD-Net dynamically represents the key SW coefficients in a multidimensional state space, thereby reducing the representation errors caused by static coefficients in traditional methods.

#### 3.2.3 Coupling residual reconstruction and learning

In the generalized SW algorithm, the surface-related correction terms jointly couple the effects of emissivity magnitude $(1 - \bar{\varepsilon})$, emissivity difference $\Delta\varepsilon$, and atmospheric water vapor content $\omega$, as follows:

$$(c_3 + c_4\omega)(1 - \bar{\varepsilon}) + (c_5 + c_6\omega)\Delta\varepsilon \tag{9}$$

This term essentially reflects the integrated residual effect of land surface and atmospheric coupling. Directly regressing the four coefficients $c_3$, $c_4$, $c_5$, and $c_6$ not only increases the number of parameters to be fitted, but also constrains the complex surface–atmosphere coupling process within a rigid parametric form, making it difficult to fully characterize higher-order nonlinear interactions under complex scenarios. To enhance the expressive flexibility of the model and simplify the parameter structure, this study rewrites the above combination as a single residual term $a_3$:

$$a_3 = (c_3 + c_4\omega)(1 - \bar{\varepsilon}) + (c_5 + c_6\omega)\Delta\varepsilon \tag{10}$$

In PCD-Net, the fourth subnetwork explicitly learns this residual term:

$$a_3 = f_{subnet,4}(\omega, \bar{\varepsilon}, \Delta\varepsilon) \tag{11}$$

With this design, PCD-Net avoids estimating multiple empirical coefficients in the emissivity correction terms separately. Instead, it treats them as an integrated compensation term jointly driven by atmospheric water vapor content, mean emissivity, and emissivity difference. This simplifies the parameter structure and provides greater expressive flexibility for characterizing complex surface–atmosphere coupling relationships. Consequently, the key physical contributions in the traditional SW formulation are mapped into four parallel learning branches: the constant-term branch, first-order brightness temperature difference branch, second-order brightness temperature difference branch, and coupling residual branch.

During the initialization phase of the parallel subnetworks, this study first fits the reference SW component parameters under various atmospheric water vapor and surface emissivity states based on the radiative transfer simulated samples. These fitted parameters serve as individual training labels for their respective branches, thereby enabling the constant, first-order, second-order, and coupling residual branches to acquire an initial representational capability that is highly consistent with their corresponding physical implications.

### 3.2.4 Two-stage fine-tuning with simulation and in situ supervision

Upon completing the initialization of the parallel subnetworks, PCD-Net seamlessly embeds the dynamically generated coefficients into the predefined SW physical structure to obtain the predicted LST:

$$T_s^{pred} = T_{10} + c_1(T_{10} - T_{11}) + c_2(T_{10} - T_{11})^2 + c_0 + a_3 \tag{12}$$

where the brightness temperature terms and the constant term correspond to the main brightness temperature response components in the classical SW structure, while $a_3$ compensates for the integrated residual jointly induced by land surface emissivity and atmospheric water vapor. To enable the four parallel branches to form a coordinated representation within the complete SW backbone, PCD-Net is first fine-tuned end-to-end in the simulation sample space. In this stage, the reference LST obtained from radiative transfer simulations, denoted as $T_s^{RTE}$, is used as the supervision label, and the mean squared error loss is defined as:

$$Loss_{sim} = \frac{1}{N_{sim}} \sum_{k=1}^{N_{sim}} \left(T_s^{pred}(k) - T_s^{RTE}(k)\right)^2 \tag{13}$$

where $N_{sim}$ is the number of simulation training samples. On this basis, site samples are further used to fine-tune the model end-to-end, thereby enhancing its adaptability to real observational conditions. Specifically, the site-derived reference LST, denoted as $T_s^{site}$, is used as the ground-truth label, and the mean squared error loss is defined as

$$Loss_{site} = \frac{1}{N_{site}} \sum_{k=1}^{N_{site}} \left( T_s^{pred}(k) - T_s^{site}(k) \right)^2 \tag{14}$$

where $N_{site}$ is the number of site samples used for fine-tuning. Different from simulation-based fine-tuning, site-supervised fine-tuning directly targets real remote sensing observational conditions. It updates the model parameters using in situ LST measurements, thereby correcting the discrepancy between the simulation environment and actual application scenarios.

### 3.2.5 LST prediction

In the prediction stage, the trained PCD-Net takes remote sensing observations and auxiliary variables as inputs and outputs the final LST retrieval result:

$$\hat{T}_s = \mathcal{F}_{PCD-Net}(T_{10}, T_{11}, \omega, \bar{\varepsilon}, \Delta\varepsilon) \tag{15}$$

To ensure model robustness and reduce the risk of overfitting, this study systematically evaluates different combinations of hidden-layer depth and neuron number through grid search (Fig. 3a) and determines the final network configuration accordingly.

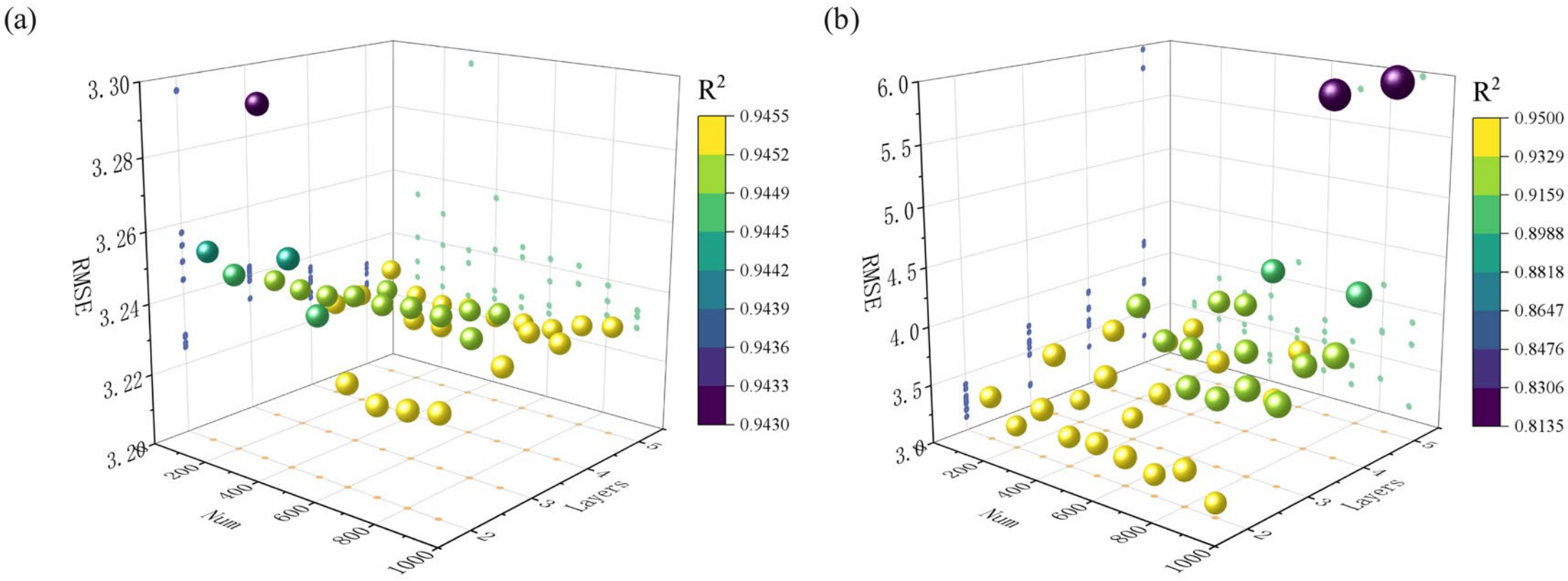

Fig. 3. Performance comparison of PCD-Net and ML under different network configurations. Different combinations of the number of hidden layers, ranging from 2 to 5 with an interval of 1, and the number of neurons, ranging from 100 to 900 with an interval of 100, were evaluated for deep neural networks. The symbol size represents the mean absolute error (MAE), the z-axis denotes the root mean square error (RMSE), and the color indicates the coefficient of determination ($R^2$). (a) PCD-Net. (b) ML.

## 3.3 Comparison methods and evaluation strategy

To systematically evaluate the performance of the proposed PCD-Net, this study uses both traditional mechanism models and data-driven models as comparison methods. Multi level validation is conducted on both the simulation dataset and the global site observation samples. The former is used to examine the theoretical fitting capability of different models under ideal radiative transfer conditions, whereas the latter is used to evaluate retrieval accuracy, stability, and generalization capability under real remote sensing observational conditions.

Two representative types of baseline models are selected. The first type includes traditional mechanism models, namely the radiative-transfer-based mechanism model and the generalized split-window-based mechanism model MM(SW). In this study, the USGS Landsat Collection 2 Level-2 LST product was used as a representative radiative-transfer-based benchmark and is denoted as MM(RT). This product is generated within a radiative transfer framework by integrating brightness temperature, land surface emissivity, vegetation indices, and reanalysis atmospheric profiles (Cook et al., 2014). MM(SW) uses the differential atmospheric absorption response of adjacent thermal infrared window bands and reduces atmospheric effects through parameterized brightness temperature combinations. It is one of the most widely used traditional methods for medium- to high-resolution thermal infrared LST retrieval. The second type is the data-driven machine learning model, denoted as ML. This model adopts a deep neural network and uses the brightness temperatures of two thermal infrared bands, $T_{10}$ and $T_{11}$, the corresponding band emissivities, $\varepsilon_{10}$ and $\varepsilon_{11}$, and atmospheric water vapor content $\omega$ as inputs to directly output LST. To ensure a fair comparison, ML and PCD-Net use information equivalent input variables and the same training procedure. Specifically, both models are first pretrained using radiative transfer simulation samples and then fine-tuned in a supervised manner using global site observation samples. The optimal ML network configuration is determined through multiple rounds of screening, as shown in Fig. 3b.

In the simulated dataset validation, this study constructs samples based on global atmospheric profiles and MODTRAN radiative transfer simulations. A 10-fold cross-validation is used to evaluate the retrieval accuracy of different methods under simulated conditions, thereby examining their ability to fit the radiative transfer relationship and the SW parameter structure. This validation mainly reflects the potential performance upper bound of different models under conditions where input variable quality is controlled. In the real world site validation, site matched samples are constructed using global ground observations, Landsat thermal infrared images, and synchronous auxiliary variables. Based on 4,450 clear-sky Landsat in-situ matchup samples from 29 sites, a site-based leave-one-out cross-validation (LOOCV) is adopted. In each iteration, one site is held out as an independent evaluation site, while samples from the remaining sites are used for model training and fine-tuning. This strategy provides a stricter test of model generalization under spatial heterogeneity and out-of-distribution conditions, and therefore more realistically reflects the application potential of the model across different global regions.

## 4. Result

### 4.1 Simulation-based theoretical accuracy assessment

#### 4.1.1 Overall accuracy validation

To evaluate the theoretical precision of each model under ideal physical conditions, a 10-fold cross-validation was conducted based on the simulated dataset (Fig. 4). It should be noted that PCD-Net$_{sim}$ in this section refers to the model after simulation-based pretraining and simulation data fine-tuning, rather than the final PCD-Net further fine-tuned using site observations. This setting excludes real site samples from training, allowing the comparison to focus more directly on the model's ability to represent the radiative transfer relationship and the SW parameter structure. Because the reference LST in the simulation dataset was derived through the closed-form solution of the RTE, MM(RT) was not included as an independent comparison method at this validation level to avoid overlap with the construction of the reference labels.

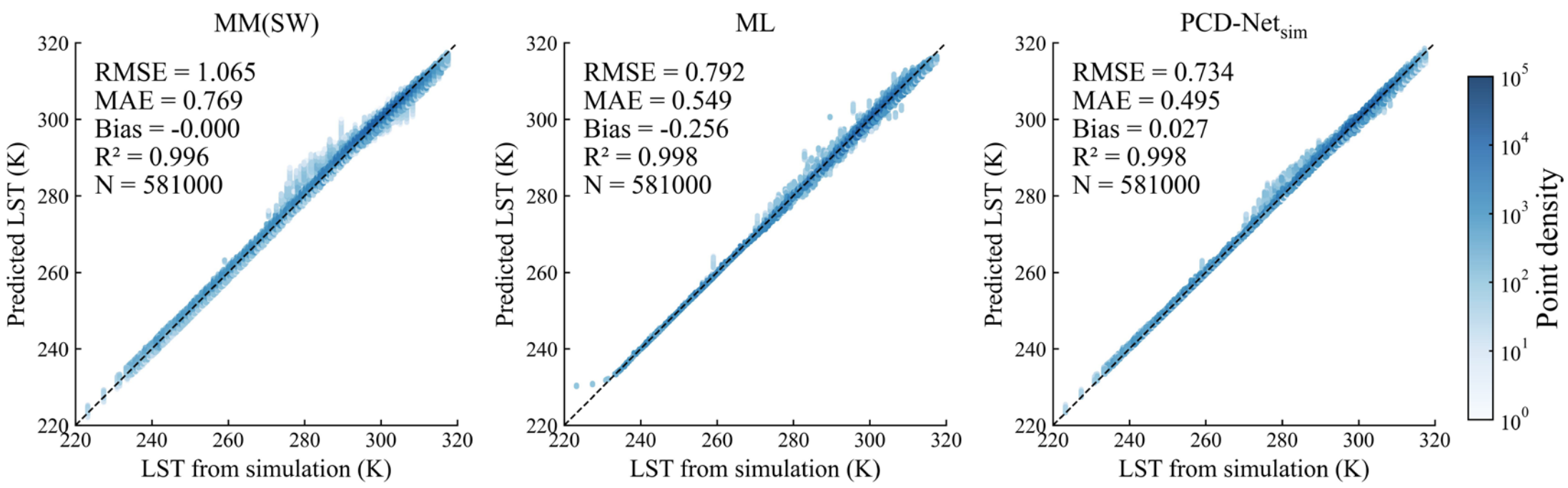

Fig. 4. Scatter plots of simulated vs. predicted LST based on 10-fold cross-validation, comparing the fitting performance of MM(SW), ML, and PCD-Net$_{sim}$ on the simulated dataset. The dashed line denotes the 1:1 line.

The results show that all three models achieved high accuracy on the simulation dataset, but clear differences remained. The traditional mechanistic model MM(SW) produced a relatively scattered distribution due to its fixed parameterized formulation, with an RMSE of 1.065 K and an MAE of 0.769 K. The ML model showed stronger nonlinear fitting capability, reducing the RMSE to 0.792 K and the MAE to 0.549 K, while increasing $R^2$ to 0.998. However, a small number of anomalous estimates still deviated from the 1:1 line in the low-temperature range below 240 K, indicating that models without an explicit physical structure may still have limited stability under boundary conditions. In contrast, PCD-Net$_{sim}$ achieved the best accuracy, with the RMSE and MAE further reduced to 0.734 K and 0.495 K, respectively. In terms of scatter distribution, the high-density central line of PCD-Net$_{sim}$ almost overlapped with the 1:1 line, and consistently high prediction agreement was maintained across the full range of the simulation dataset.

#### 4.1.2 Error distributions across LST and water vapor conditions

To further evaluate model stability under different LST and atmospheric conditions, the RMSE distributions of different methods were calculated across different LST and atmospheric water vapor content intervals, as shown in Fig. 5. Overall, the errors of all three methods increased with increasing LST and water vapor content, reaching relatively high levels under high-temperature and high-humidity conditions. This indicates that complex thermal states and strong atmospheric absorption substantially increase the uncertainty of LST retrieval.

Clear differences were observed in the error responses of different methods. MM(SW) showed a more rapid error increase in high water vapor intervals, indicating the limited adaptability of the fixed-parameter SW model under strong water vapor conditions. ML achieved lower overall errors than MM(SW), but local fluctuations occurred at the low temperature, suggesting that models without an explicit physical structure may still produce unstable estimates under boundary conditions. In comparison, PCD-Net$_{sim}$ maintained lower RMSE values in most intervals, with fewer high-error regions.

Notably, the error distribution pattern of PCD-Net$_{sim}$ was more similar to that of MM(SW), both showing a smooth error structure that changed continuously with environmental variables. By contrast, the error distribution of ML was more prone to local discrete fluctuations. This result indicates that PCD-Net$_{sim}$ retains the structural stability provided by the SW physical backbone while enhancing nonlinear representation capability. Therefore, the proposed model does not simply replace the mechanism model but adaptively corrects complex coupling errors based on the SW parameter structure.

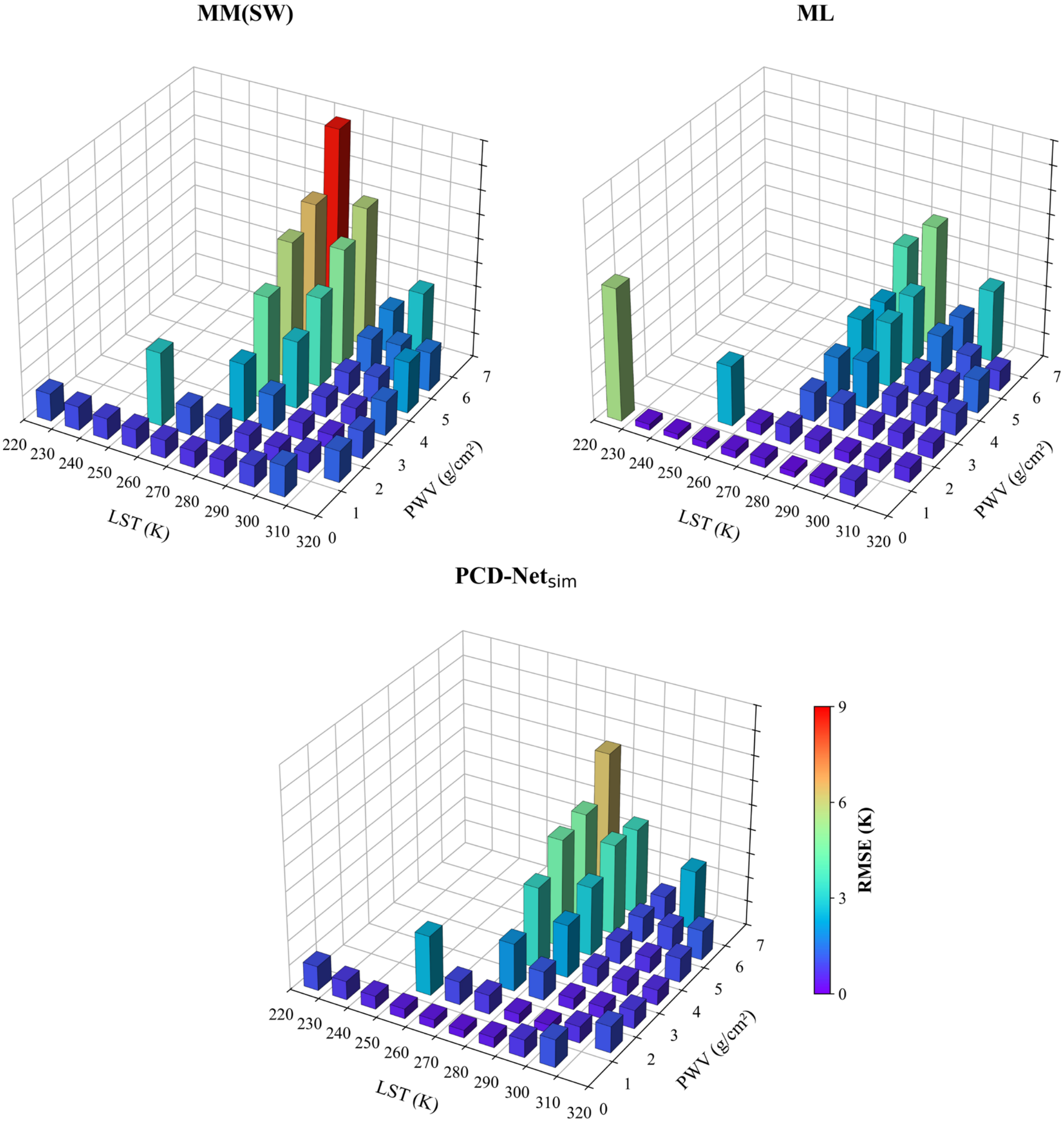

Fig. 5. RMSE distributions of MM(SW), ML, and PCD-Net$_{sim}$ across different LST and atmospheric water vapor content intervals. The color represents the RMSE value within each interval.

#### 4.1.3 Sensitivity analysis of key input variables

In LST retrieval, uncertainties in input variables directly affect the stability and accuracy of retrieval results. Therefore, this study selected at-sensor radiance in the thermal infrared bands, land surface emissivity, and atmospheric water vapor content as the main uncertainty factors. Correlation analysis was first conducted to reveal the statistical relationships between each input variable and LST, followed by sensitivity experiments.

The correlation analysis results (Fig. 6) show that LST has a very strong linear relationship with at-sensor radiance in the thermal infrared bands. The correlation coefficients of $L_{10}$ and $L_{11}$ reached 0.96 and 0.95, respectively, indicating that at-sensor radiance was the variable most strongly correlated with LST in the simulation dataset. Atmospheric water vapor content showed a moderate positive correlation with LST, with a correlation coefficient of 0.51, suggesting a moderate association between water vapor and LST in the simulation dataset. In contrast, land surface emissivity showed weak direct linear correlations with LST, indicating that emissivity itself does not mainly act as a univariate linear control on LST. However, in SW retrieval, emissivity participates in error compensation through $(1 - \bar{\varepsilon})$, $\Delta\varepsilon$, and their coupling with water vapor content. It is therefore an important modulating factor in the conversion from brightness temperature to LST and should be further evaluated in the following sensitivity experiments.

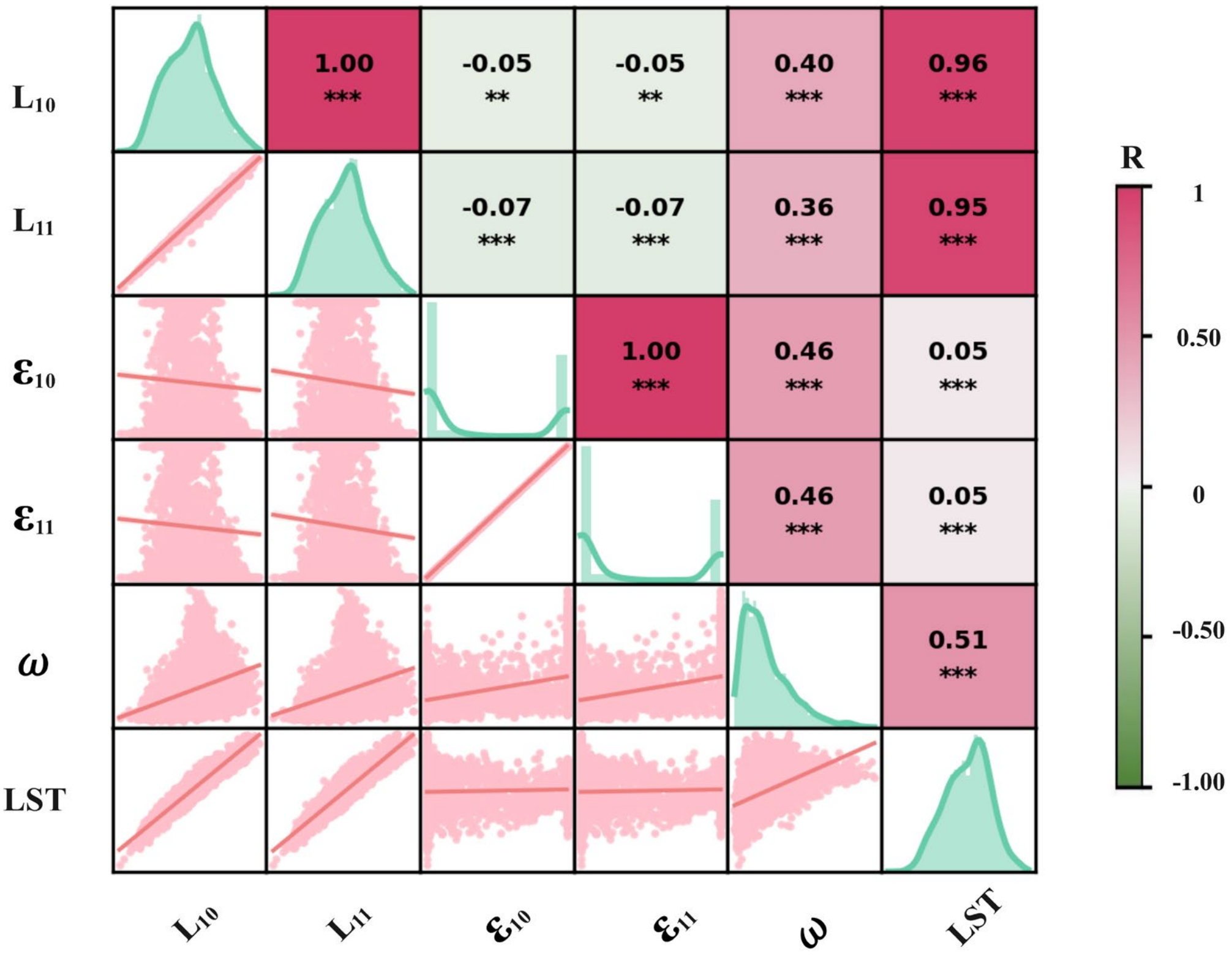

Fig. 6 Correlation matrix between input variables $L_{10}$, $L_{11}$, $\varepsilon_{10}$, $\varepsilon_{11}$, $\omega$, and LST. *** and ** indicate statistical significance at $p < 0.001$ and $p < 0.01$, respectively.

Under land surface emissivity perturbations (Fig. 7a), the mean LST errors of MM(SW) and PCD-Net$_{sim}$ varied only slightly and were generally within ±0.4 K. Although the mean error curve of MM(SW) was the flattest, PCD-Net$_{sim}$ showed a narrower standard deviation band, indicating lower error dispersion under emissivity perturbations. For at-sensor radiance perturbations (Fig. 7b), the errors of all three methods changed approximately linearly with radiance bias, with maximum deviations exceeding ±4 K. This indicates that, within the perturbation range considered in this study, at-sensor radiance error was the most significant source of uncertainty in LST retrieval. Under this highly sensitive condition, the mean error curve of PCD-Net$_{sim}$ was closer to the zero-error line, and its standard deviation band was relatively narrow, suggesting a more stable error response to radiance perturbations. For atmospheric water vapor content perturbations (Fig. 7c), the mean error curve and error variation range of PCD-Net$_{sim}$ were generally closer to the zero-error line than those of ML, indicating smaller systematic bias and a more stable error response under water vapor perturbations.

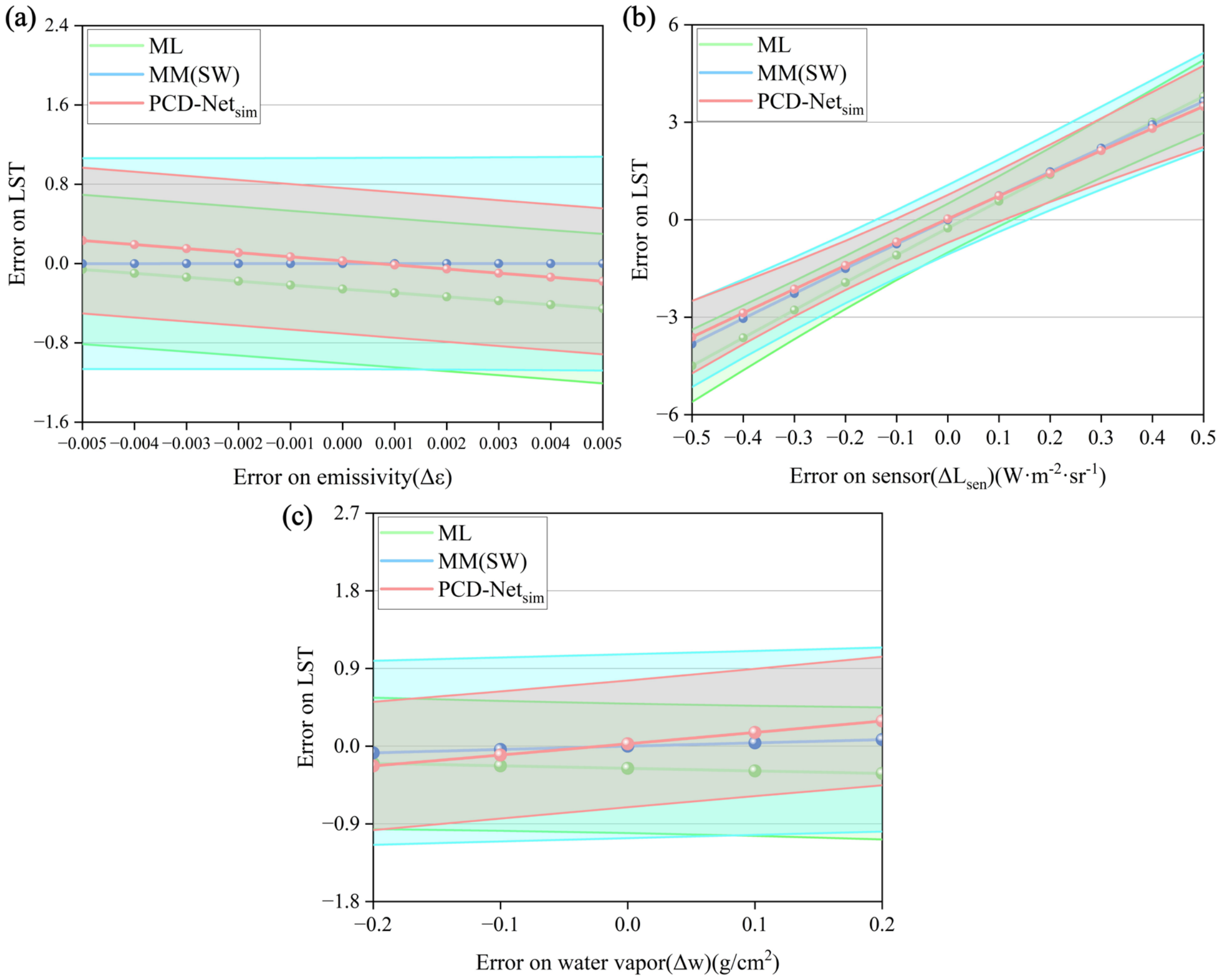

Fig. 7. Sensitivity analysis of LST retrieval errors to input variable perturbations. (a) Land surface emissivity perturbation; (b) at-sensor radiance perturbation; (c) atmospheric water vapor content perturbation. The green, blue, and red curves represent the mean errors of ML, MM(SW), and PCD-Net$_{sim}$, respectively, and the shaded areas indicate the corresponding standard deviations.

## 4.2 Global site validation

### 4.2.1 Overall site validation results

The global site validation results show clear differences in retrieval accuracy among different methods under real remote sensing observational conditions (Fig. 8). The traditional mechanism models MM(SW) and MM(RT) yielded RMSE values of 3.78 K and 4.33 K, MAE values of 2.81 K and 3.30 K, and $R^2$ values of 0.925 and 0.902, respectively. These results indicate that, under globally complex surface and atmospheric conditions, both the traditional SW parameterization method and the MM(RT) relying on auxiliary atmospheric parameters still face substantial uncertainties.

For the data-driven methods, two machine learning models were further distinguished in this study. The pure ML (ML$_{pure}$) was trained only with site samples, without simulation-sample training, and achieved an RMSE of 3.55 K, an MAE of 2.75 K, and an $R^2$ of 0.934. In comparison, the ML model trained with simulation samples and further fine-tuned using site observations reduced the RMSE to 3.12 K and the MAE to 2.34 K, while increasing $R^2$ to 0.949. This indicates that prior information provided by simulation samples helps improve the parameter learning capability of data-driven models under real site conditions.

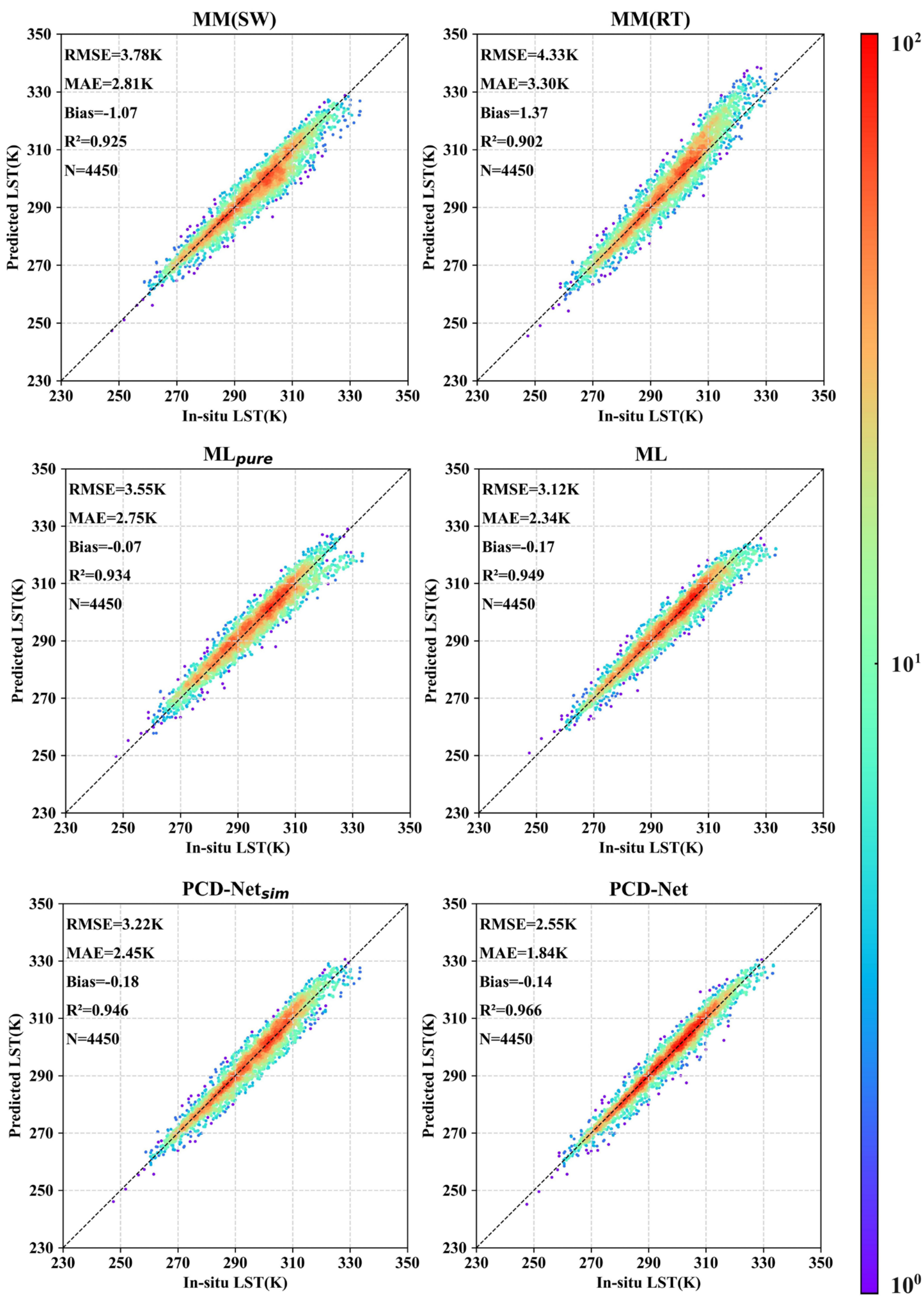

Fig. 8. LST validation results of six models at 29 global ground sites, including MM(SW), MM(RT), $ML_{pure}$, ML, $PCD\text{-}Net_{sim}$, and PCD-Net. The color represents scatter density, and the dashed line denotes the 1:1 line.

$PCD\text{-}Net_{sim}$ achieved an RMSE of 3.22 K, an MAE of 2.45 K, and an $R^2$ of 0.946, outperforming the traditional mechanism models and $ML_{pure}$ trained only with site samples. This indicates that the SW physical backbone and component decoupled structure have good transfer potential from simulated conditions to real observational scenarios. Nevertheless, $PCD\text{-}Net_{sim}$ still performed slightly worse than the site-supervised ML model, suggesting that non-negligible distribution discrepancies exist between simulation samples and real site observations. Therefore, site-supervised information is necessary for correcting systematic biases under real remote sensing observational conditions.

After further introducing site-supervised fine-tuning, the final PCD-Net achieved the best overall accuracy, with the RMSE and MAE reduced to 2.55 K and 1.84 K, respectively, and $R^2$ increased to 0.966. Compared with ML using the same simulation-sample training and site-supervised fine-tuning procedure, PCD-Net still showed clear advantages. This demonstrates that its accuracy improvement is not only attributable to site-supervised adaptation, but also to the additional structural gains provided by the SW physical backbone and the parallel component decoupled architecture.

These comparisons clarify the respective roles of $ML_{pure}$, ML, and PCD-$Net_{sim}$. $ML_{pure}$ is retained to isolate the effect of simulation derived priors, whereas ML adopts the same simulation pretraining and site-supervised fine-tuning procedure as PCD-Net but does not include the SW physical backbone or the parallel component decoupled architecture. Therefore, the comparison between ML and PCD-Net more directly reflects the additional benefit of the proposed structural design. PCD-$Net_{sim}$ is used as an intermediate model to examine the transferability of the simulation-trained PCD-Net before site-supervised fine-tuning. Accordingly, the following site level, spatial, temporal, and extreme-condition analyses mainly focus on the comparison between ML and the final site-fine-tuned PCD-Net.

#### 4.2.2 Site level performance across regions and climates

The site level MAE results further reveal the performance differences of different methods across regions and climatic conditions (Fig. 9). MM(SW) maintained relatively stable accuracy at some mid- and high-latitude sites. For example, the MAE values at high-latitude sites such as BAR, NYA, and CAP were approximately 2 K, while those at mid latitude sites such as FPK, SXF, DM, SDQ, and ZYSD were below 2 K. However, MM(SW) showed substantially increased errors under high-temperature and high-humidity environments, such as the tropical site OHY with an MAE of 6.96 K and the humid high water vapor site TAT with an MAE of 6.60 K.

MM(RT) showed larger biases at hot and arid sites, such as GOB with an MAE of 5.72 K and OHY with an MAE of 4.98 K, indicating that existing radiative-transfer-based products still have clear limitations in complex hot–dry environments. In contrast, ML achieved lower overall errors, with a global MAE of 2.34 K, suggesting that simulation initialization and site-supervised fine-tuning can effectively improve the retrieval accuracy of data-driven models under real observational conditions. Nevertheless, ML still showed relatively high errors at boundary condition sites such as OHY and TAT.

PCD-Net achieved the lowest or near-lowest MAE at most sites, with site level MAE values mainly concentrated between 1.4 K and 2.5 K, indicating good spatial stability. In particular, at high latitude cold region sites such as NYA and CAP, as well as typical arid region sites such as GOB and IZA, the MAE values of PCD-Net remained below 2 K.

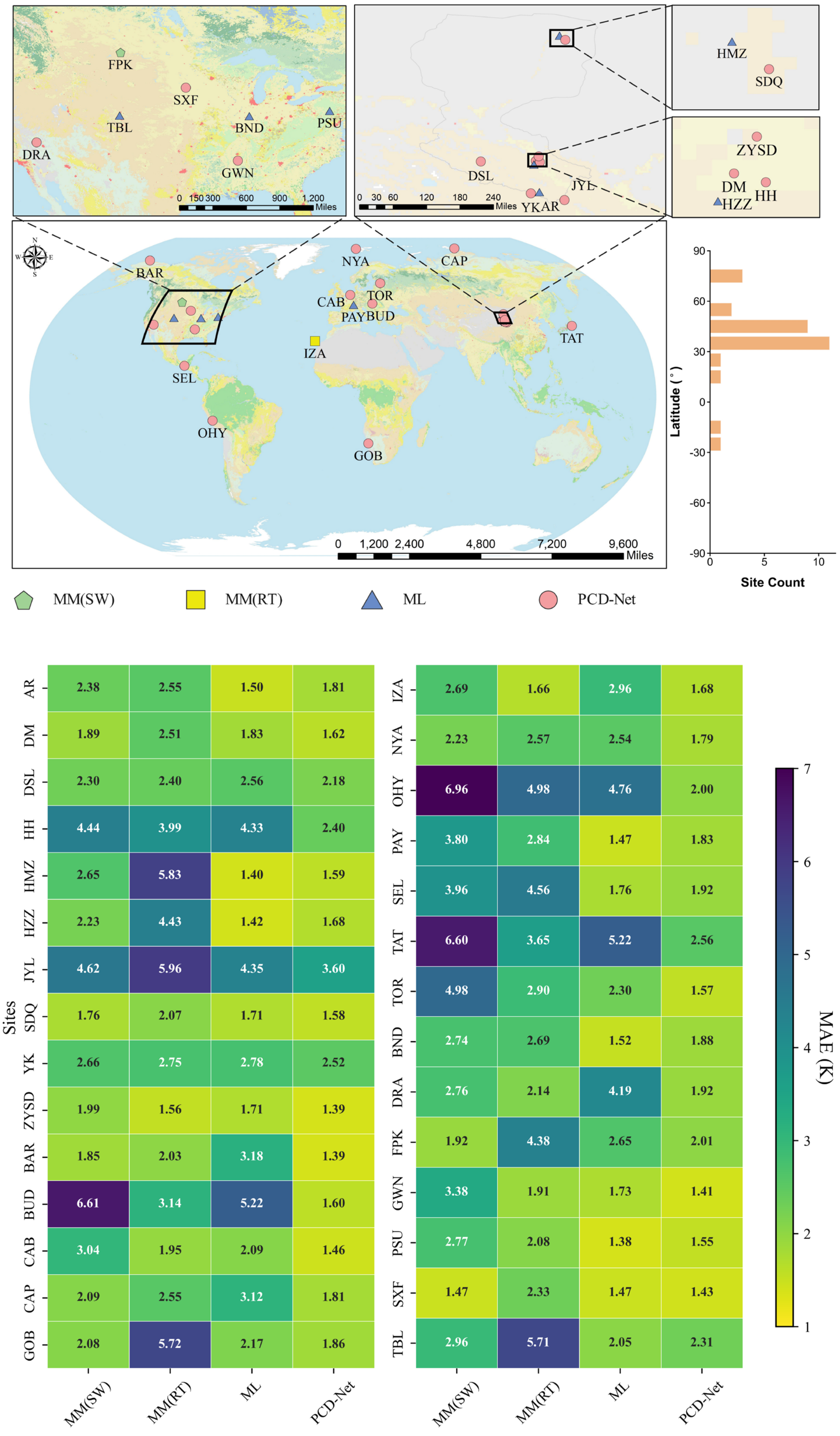
FPK
SXF
TBL
BND
PSU
DRA
GWN
0 150 300 600 900 1,200 Miles
HMZ
SDQ
ZYSD
DM
HH
HZZ
DSL
JYL
YKAR
0 30 60 120 180 240 Miles
BAR
NYA
CAP
CAB
TOR
PAYBUD
TAT
IZA
SEL
OHY
GOB
0 1,200 2,400 4,800 7,200 9,600 Miles
Latitude (°)
Site Count
MM(SW)
MM(RT)
ML
PCD-Net
Sites
AR 2.38 2.55 1.50 1.81
DM 1.89 2.51 1.83 1.62
DSL 2.30 2.40 2.56 2.18
HH 4.44 3.99 4.33 2.40
HMZ 2.65 5.83 1.40 1.59
HZZ 2.23 4.43 1.42 1.68
JYL 4.62 5.96 4.35 3.60
SDQ 1.76 2.07 1.71 1.58
YK 2.66 2.75 2.78 2.52
ZYSD 1.99 1.56 1.71 1.39
BAR 1.85 2.03 3.18 1.39
BUD 6.61 3.14 5.22 1.60
CAB 3.04 1.95 2.09 1.46
CAP 2.09 2.55 3.12 1.81
GOB 2.08 5.72 2.17 1.86
IZA 2.69 1.66 2.96 1.68
NYA 2.23 2.57 2.54 1.79
OHY 6.96 4.98 4.76 2.00
PAY 3.80 2.84 1.47 1.83
SEL 3.96 4.56 1.76 1.92
TAT 6.60 3.65 5.22 2.56
TOR 4.98 2.90 2.30 1.57
BND 2.74 2.69 1.52 1.88
DRA 2.76 2.14 4.19 1.92
FPK 1.92 4.38 2.65 2.01
GWN 3.38 1.91 1.73 1.41
PSU 2.77 2.08 1.38 1.55
SXF 1.47 2.33 1.47 1.43
TBL 2.96 5.71 2.05 2.31
MM(SW) MM(RT) ML PCD-Net
MAE (K)

Fig. 9. MAE distributions of four models across 29 global ground observation sites on five continents. The upper panel shows the method with the lowest MAE at each site among the four main comparison methods, with MM(SW), MM(RT), ML, and PCD-Net represented by green pentagons, yellow squares, blue triangles, and red circles, respectively. The lower panel shows the site level MAE values of the four models. Detailed site information is provided in Table 1.

### 4.2.3 Regional spatial comparison and time-series validation

To further evaluate the spatial continuity and the ability of different methods to represent land surface thermal heterogeneity at the regional scale, four representative regions in Asia (AR), Europe (CAB), Africa (IZA), and North America (TBL) were selected. The retrieval results of MM(SW), MM(RT), ML, and PCD-Net were spatially compared with the MODIS LST product, hereafter referred to as MOD LST (Fig. 10). Because MOD LST differs from the Landsat-based results in spatial resolution and valid-pixel coverage, it was used only as a reference for regional temperature patterns rather than as pixel-level ground truth.

Overall, traditional mechanism models and the ML method showed evident spatial biases in different regions. MM(RT) produced localized high-temperature enhancement in several regions, especially over bare soil, arid, or sparsely vegetated surfaces. ML captured certain spatial temperature differences, but it also produced excessively strong local high-temperature responses and unnatural spatial discontinuities in some regions, indicating its sensitivity to input perturbations and changes in sample distribution. By contrast, PCD-Net generally showed more continuous spatial transitions and more stable temperature distributions in the four representative regions. It preserved land surface thermal heterogeneity while suppressing abnormal high values and local discontinuities.

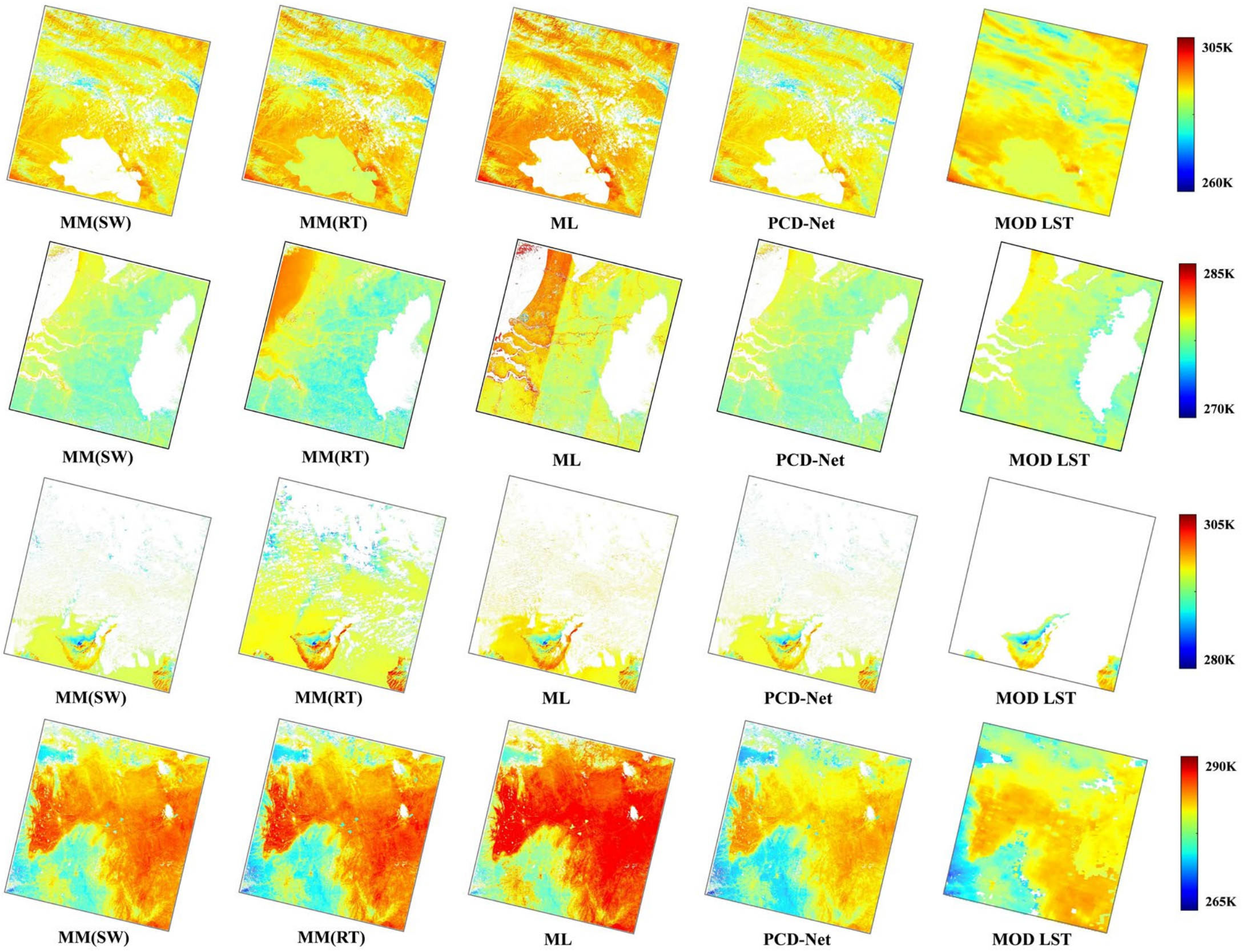

Fig. 10. Spatial comparison of LST distributions retrieved by MM(SW), MM(RT), ML, and PCD-Net in four representative site-centered regions, with the MODIS LST product (MOD LST) used as a reference. The four regions

are centered on the AR site in Asia, the CAB site in Europe, the IZA site in Africa, and the TBL site in North America, respectively, and show the temperature spatial patterns within the surrounding Landsat 8 coverage. White areas indicate invalid or missing pixels.

Specifically, in the AR region of Asia, both MM(RT) and ML showed evident high-temperature enhancement, with substantially higher temperatures in some local areas. PCD-Net produced a more balanced temperature distribution, preserving the main spatial gradient while reducing the excessive amplification of high value regions. In the European region, ML produced obvious striped or blocky high-temperature anomalies, while PCD-Net showed a more continuous spatial distribution and an overall pattern closer to the MOD LST reference. In the arid region of Africa, the spatial comparison was partly limited by the small number of valid MOD LST pixels. Nevertheless, within the valid coverage, PCD-Net preserved the main high-temperature structure while avoiding the pronounced local high-temperature expansion observed in ML. In the North American region, ML showed large-scale overestimation, and MM(SW) and MM(RT) also showed different degrees of positive bias. PCD-Net was more consistent with the large-scale temperature pattern reflected by MOD LST in terms of both overall temperature level and spatial distribution, indicating better regional mapping stability.

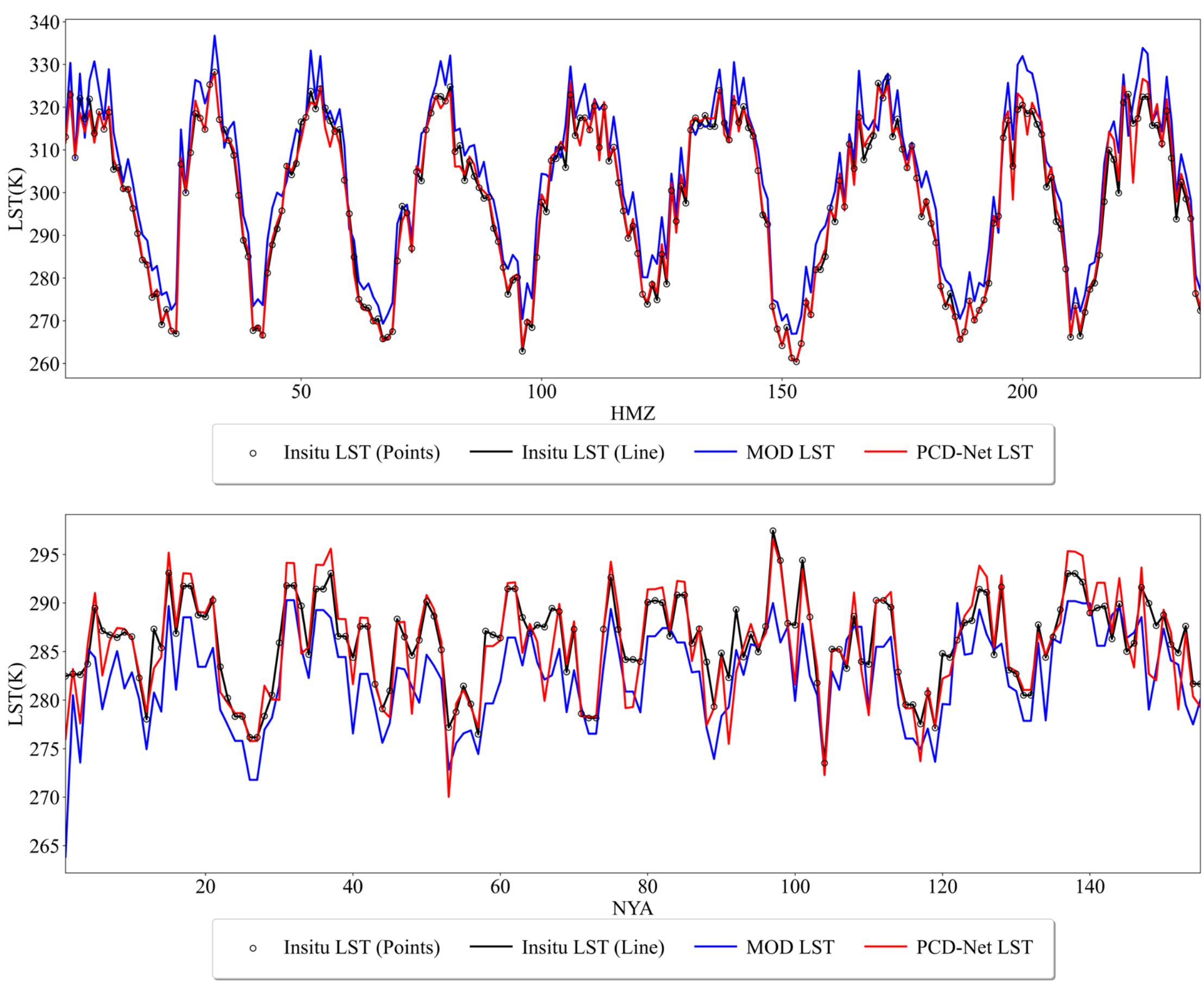

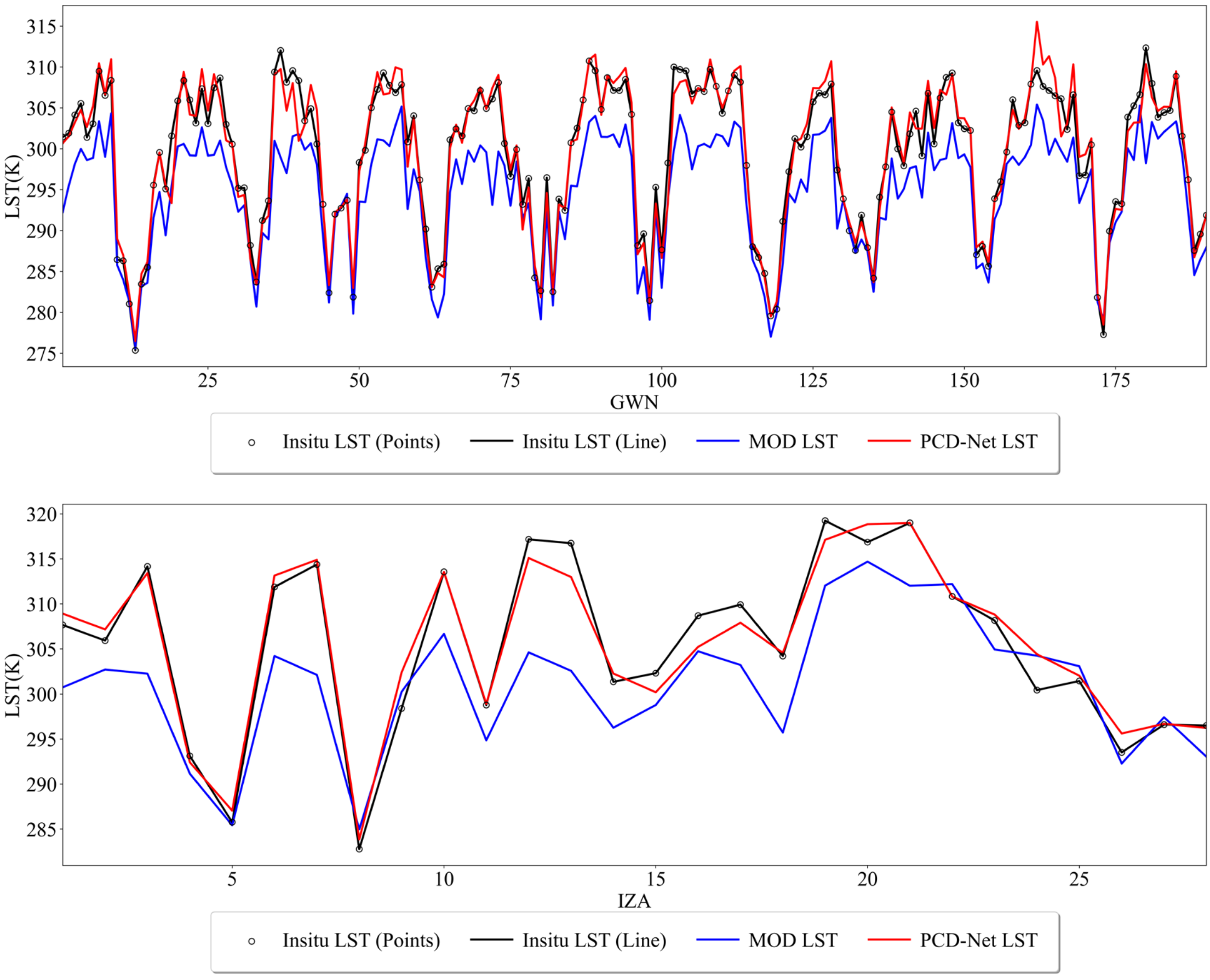

Fig. 11. Long term time-series comparison among PCD-Net, the MODIS LST product, and ground observations at four representative sites. The four sites are located in Asia (HMZ), Europe (NYA), North America (GWN), and Africa (IZA), respectively.

To further evaluate the ability of PCD-Net to characterize temporal variations in LST, four representative sites in Asia (HMZ), Europe (NYA), North America (GWN), and Africa (IZA) were selected for long term time-series comparison among PCD-Net retrievals, MOD LST, and in situ observations (Fig. 11). Overall, PCD-Net effectively tracked temporal variations in LST at different sites and was closer to ground observations than the MODIS product in most cases.

Specifically, the HMZ site in Asia exhibited relatively complex temporal fluctuations. MOD LST showed substantial deviations, while PCD-Net better fitted the observed time series, especially during high-temperature periods. At the NYA site in Europe, MOD LST showed large errors and pronounced fluctuations, while PCD-Net stably tracked the observed sequence and substantially narrowed the error range. At the GWN site in North America and the IZA site in Africa, PCD-Net agreed well with in situ LST, capturing both long term trends and short-term fluctuations. MOD LST, by contrast, showed systematic underestimation at both sites.

### 4.3 Extreme condition performance

The quantitative analysis results (Fig. 12) show that the error distributions of different methods differed substantially under extreme LST and atmospheric water vapor conditions. Under extremely high-temperature and extremely high water vapor conditions, the errors of traditional mechanism models were markedly amplified. For MM(RT), the median absolute error exceeded 4 K under the extremely high-temperature scenario, with outliers reaching approximately 14 K. This indicates that MM(RT) retrievals relying on auxiliary atmospheric parameters can still produce large deviations under high-temperature conditions. For MM(SW), the median absolute error increased to more than 6 K under extremely high water vapor conditions, and the interquartile range also expanded considerably, indicating insufficient stability of the fixed-parameter SW model under strong water vapor conditions.

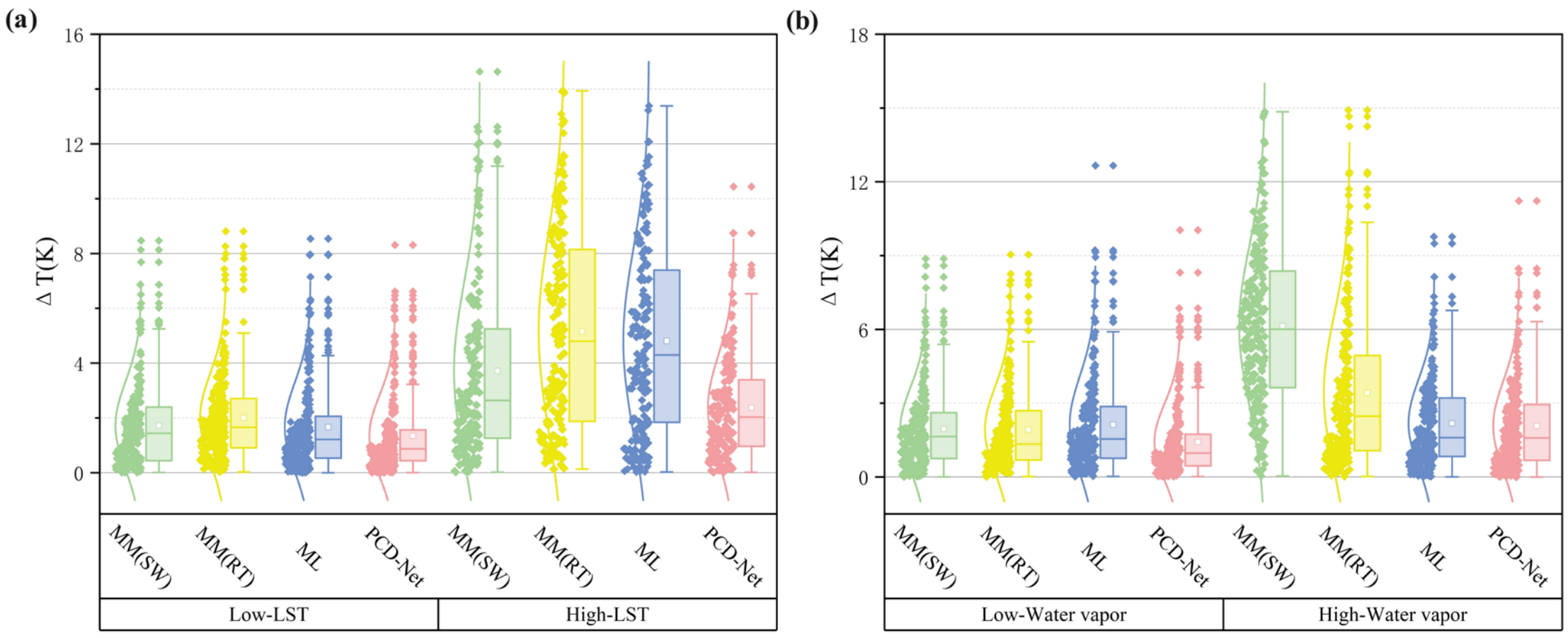

Fig. 12. Boxplot statistics of LST absolute errors under extreme LST and atmospheric water vapor conditions based on ground validation data. (a) Comparison between the lowest 5% and highest 5% LST samples. (b) Comparison between the lowest 5% and the highest 5% atmospheric water vapor content samples.

Under extremely low-temperature and extremely low water vapor conditions, the overall error levels of all methods decreased, and the differences among models became smaller. The median errors of MM(SW), MM(RT), and ML in these two scenarios were mostly around 2 K, while extreme deviations were mainly distributed above 4 K. Under the same conditions, PCD-Net further reduced both the median error and the tail errors. The median error in the extremely low temperature subset was controlled within 1 K, and the lowest overall error level was also maintained in the extremely low water vapor subset. These results indicate that PCD-Net can not only mitigate error amplification in difficult scenarios such as high-temperature and high water vapor, but also maintain a more compact error distribution under low temperature and low water vapor conditions.

The visualization results at representative extreme sites (Fig. 13) further reveal the spatial differences among different models under extreme conditions. It should be noted that MOD LST may also contain uncertainties under extreme temperature and water vapor conditions. Therefore, it was mainly used as a reference for the regional temperature background and large-scale spatial pattern, rather than as strict pixel-level ground truth. The temperature and water vapor conditions of the corresponding sites, together with the site level biases of different methods, are provided in the caption of Fig. 13.

In the extremely low-temperature scenario at the CAB site (Fig. 13a), all methods generally reflected the low-temperature background, but clear differences appeared in local spatial structures. ML showed relatively strong high value responses in some areas, with less continuous spatial transitions. MM(RT) also produced localized temperature enhancement. In comparison, PCD-Net produced a more balanced temperature distribution and more continuous spatial transitions while maintaining the low-temperature background. Its regional pattern was more consistent with the large-scale low-temperature background reflected by MOD LST.

In the extremely low water vapor scenario at the DM site (Fig. 13b), different methods showed generally similar representations of the regional temperature gradient, but local differences remained. MM(RT) and ML showed relatively strong high-temperature responses in some areas, and the local high value expansion of ML was more pronounced. The spatial patterns of MM(SW) and PCD-Net were more consistent with the regional temperature background shown by MOD LST. PCD-Net further showed more continuous temperature variations over terrain transitions and boundary areas, reducing local abrupt changes.

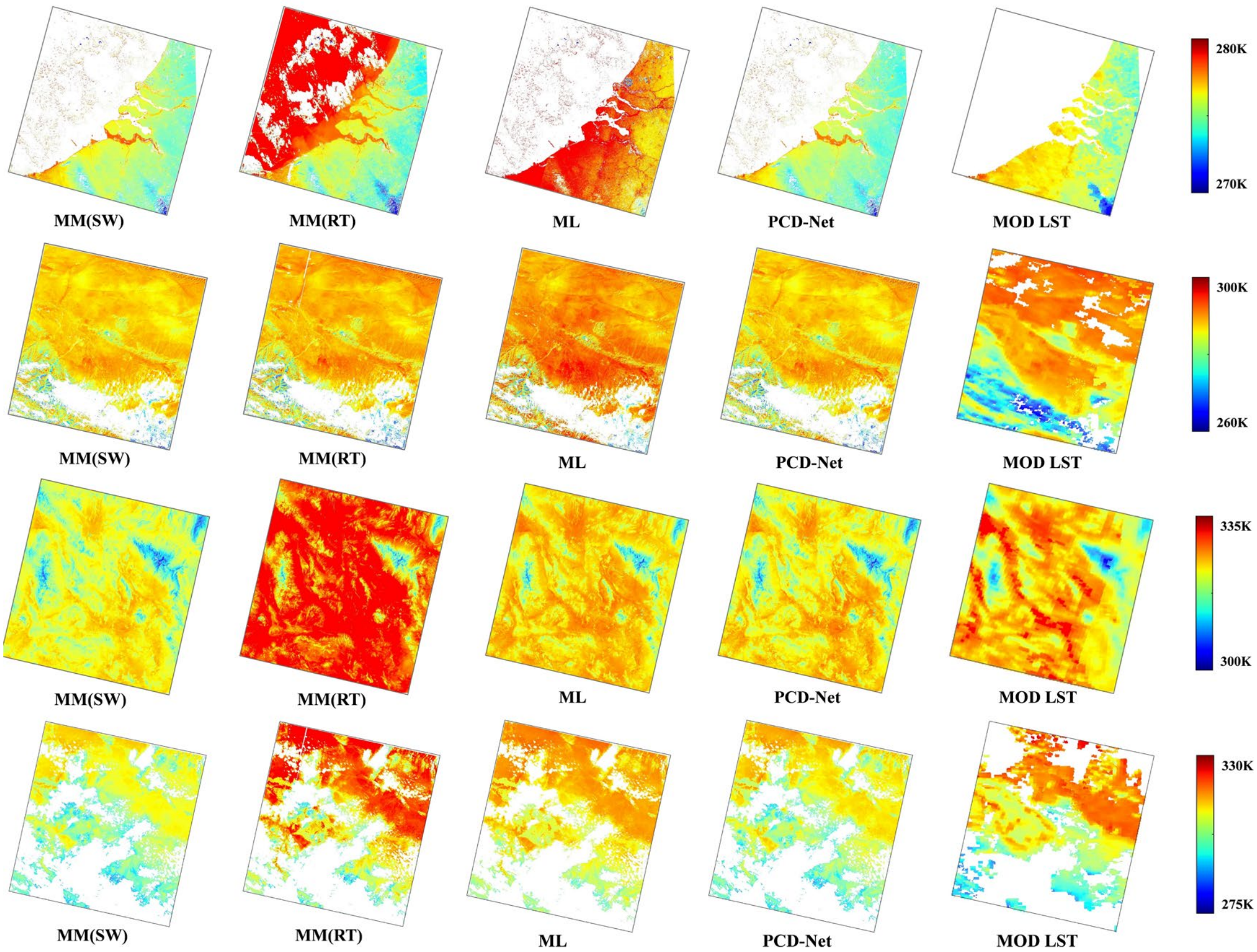

Fig. 13. Visual comparison of LST retrieval results under four extreme scenarios. The figure shows the spatial distribution differences among MM(SW), MM(RT), ML, PCD-Net, and the MOD LST reference product. Site level bias is calculated as the difference between each result and the in situ LST. The values are listed in the order of MM(SW), MM(RT), ML, PCD-Net, and MOD LST. (a) Extremely low LST at the CAB site in Europe, with an in situ LST of 274.64 K and site level biases of −0.71 K, −1.31 K, −1.72 K, −0.29 K, and −0.54 K. (b) Extremely low atmospheric water vapor content at the DM site in Asia, with $\omega$=0.08 g·cm$^{-2}$ and site level biases of 1.71 K, 1.53 K, 1.14 K, 0.59 K, and −0.08 K. (c) Extremely high LST at the DRA site in North America, with an in situ LST of 328.55 K and site level biases of −4.89 K, 5.27 K, −7.04 K, −0.91 K, and −0.98 K. (d) Extremely high atmospheric water vapor content at the HZZ site in Asia, with $\omega$=3.37 g·cm$^{-2}$ and site level biases of −6.30 K, 10.10 K, −1.98 K, −0.08 K, and −0.45 K.

In the extremely high-temperature scenario at the DRA site (Fig. 13c), the differences among models became more pronounced. MM(RT) showed large-scale high value expansion, and its temperature field differed substantially from the other results and the MOD LST reference pattern. ML preserved part of the high-temperature spatial structure, but its local high value patches were relatively fragmented, and temperature transitions were insufficiently continuous in some areas. MM(SW) captured the main high-temperature pattern, but its response in the core high-temperature region was relatively weak. In comparison, PCD-Net better preserved the main spatial structure of the high-temperature region while avoiding the excessive high value expansion observed in MM(RT).

In the extremely high water vapor scenario at the HZZ site (Fig. 13d), strong water vapor further amplified the spatial differences among different methods. MM(RT) showed obvious large-scale high value expansion, while MM(SW) produced a relatively low overall temperature level and showed weak representation of local temperature gradients. ML improved the spatial pattern in some areas, but local high value expansion and spatial discontinuities remained. In comparison, PCD-Net suppressed abnormal high value expansion while preserving the main spatial variation structure. Its overall temperature distribution was more continuous and showed better consistency with the large-scale temperature pattern indicated by MOD LST.

## 5. Discussion

### 5.1 Core architecture ablation

To evaluate the contribution of key structural designs in the proposed framework to performance improvement, this study constructed several ablation variants with different architectural and training settings. These variants include the complete PCD-Net, PCD-Net$_{npt}$ without initialization pretraining, PCD-Net$_{dir}$ that removes the coupling residual branch and directly regresses the seven SW coefficients, and PCD-Net$_{dir, npt}$ that removes both initialization pretraining and the residual branch.

**Table 2**

Ablation study of the residual branch and pretraining strategy in PCD-Net.

| Method | Residual branch | Pretraining | Simulation 10-fold | | | Site LOOCV | | |
|---|---|---|---|---|---|---|---|---|
| | | | MAE | RMSE | $R^2$ | MAE | RMSE | $R^2$ |
| PCD-Net$_{dir, npt}$ | × | × | 0.66 | 0.85 | 0.997 | 2.51 | 3.28 | 0.944 |
| PCD-Net$_{dir}$ | × | √ | 0.55 | 0.73 | 0.998 | 2.46 | 3.22 | 0.946 |
| PCD-Net$_{npt}$ | √ | × | 0.49 | 0.72 | 0.998 | 1.99 | 2.86 | 0.957 |
| PCD-Net | √ | √ | 0.49 | 0.73 | 0.998 | 1.84 | 2.55 | 0.966 |

As shown in Table 2, in the 10-fold cross-validation based on the MODTRAN simulation dataset, PCD-Net achieved an RMSE of 0.73 K and an MAE of 0.49 K, which were close to those of the more complex PCD-Net$_{dir}$. However, without initialization pretraining, the four-branch PCD-Net$_{npt}$ still maintained an RMSE of 0.72 K, whereas the more complex seven-parameter regression model PCD-Net$_{dir,npt}$ degraded to 0.85 K. This indicates that, under ideal physical conditions, the four-branch parallel architecture based on the constant term, first-order term, second-order term, and coupling residual term is sufficient to represent the main radiative transfer relationship. In contrast, the direct regression structure with higher degrees of freedom is more prone to instability.

When the validation scenario shifted to real observational conditions, the advantage of the component decoupled structure became more evident. In the site-based LOOCV, PCD-Net achieved an RMSE of 2.55 K, which was substantially lower than that of PCD-Net$_{dir}$ at 3.22 K. Meanwhile, compared with PCD-Net$_{npt}$, which had an RMSE of 2.86 K without initialization pretraining, the complete model also achieved higher accuracy. These results indicate that explicitly mapping network outputs to different physical components in the SW formulation and using the residual branch to learn surface–atmosphere coupling corrections can effectively constrain the solution space and improve model stability and retrieval capability under complex conditions.

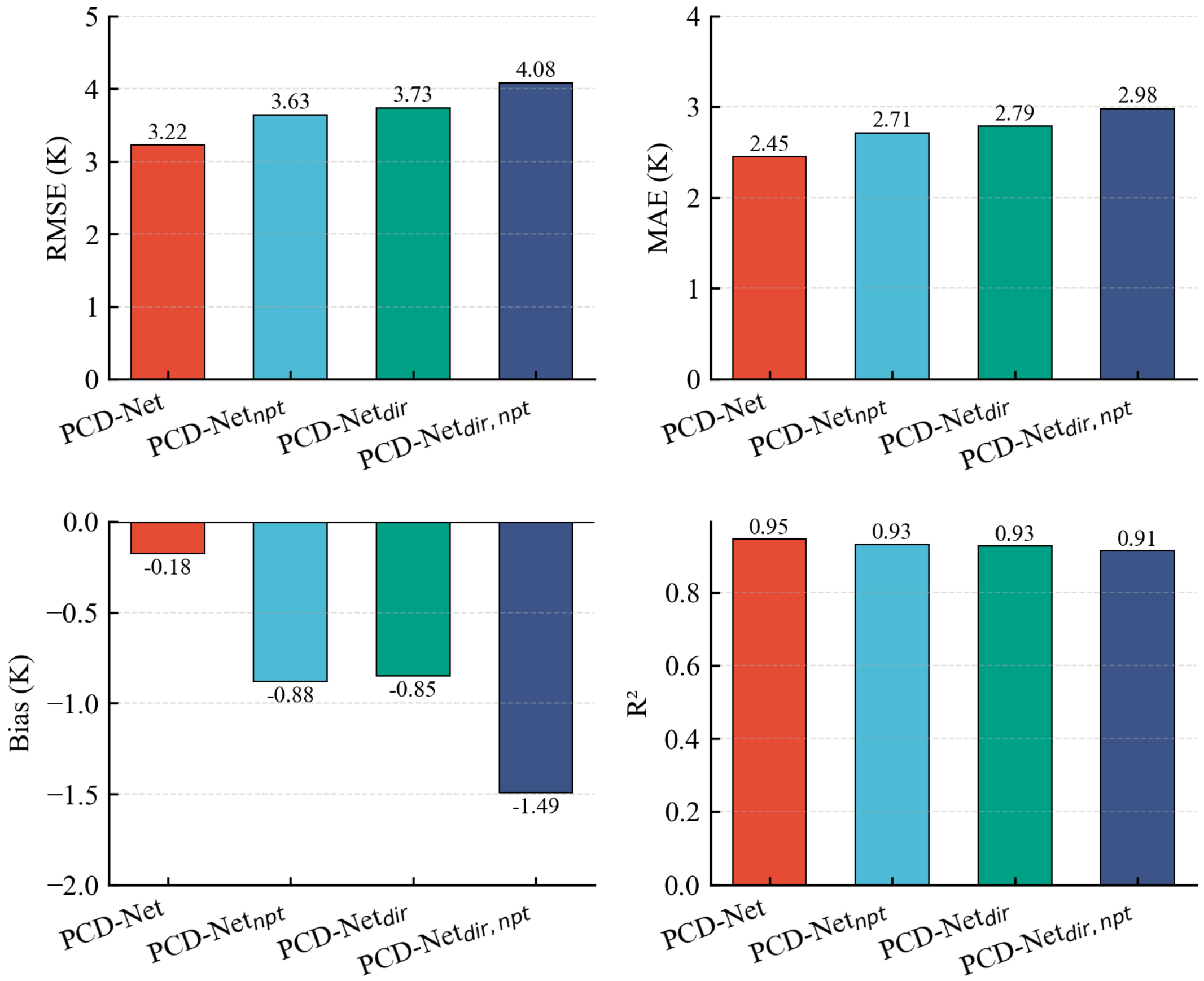

Fig. 14. Statistical comparison of retrieval accuracy metrics for PCD-Net and its ablation variants in independent site validation.

A consistent trend was also observed in the independent site validation (Fig. 14). In this validation, models were trained only using the simulation dataset and were then directly applied to real site samples without any site-supervised fine-tuning. Therefore, this experiment provides a more direct assessment of the transfer capability of different structures from simulated conditions to real observational conditions. The results show that simulation-based initialization substantially reduced the systematic bias of both architectural variants in real-site prediction. For PCD-Net, the bias converged from −0.88 K in PCD-$\text{Net}_{npt}$ to −0.18 K. For PCD-$\text{Net}_{dir}$, the bias converged from −1.49 K in PCD-$\text{Net}_{dir,npt}$ to −0.85 K. These results suggest that simulation-based initialization pretraining can provide a more reasonable parameter initialization for subsequent site-supervised fine-tuning, thereby suppressing systematic bias and enhancing model adaptability across scenarios. This finding is also consistent with the general rationale of physics-constrained deep learning and mechanism-guided data-driven modeling in recent studies (Reichstein et al., 2019).

## 5.2 Explicit physical decoupling versus multi-output learning

To further evaluate the performance advantage of the parallel subnetwork architecture over a single-backbone multi-output architecture, this study compared PCD-Net with a representative holistic multi-output network, termed M-Net. M-Net uses a single backbone network to directly regress all SW coefficients $c_0$–$c_6$ (Cheng et al., 2025). The comparison covered multiple validation levels from simulation data to real site observations, allowing the retrieval capabilities of the two architectures to be evaluated under different data conditions (Fig. 15). In the simulated 10-fold cross-validation based on the simulation dataset, both PCD-Net and M-Net achieved high fitting accuracy, with MAE values of 0.49 K and 0.54 K, respectively. This indicates that, under simulated conditions with controlled input quality and sufficient sample coverage, a holistic multi-output network can also learn the numerical mapping between radiative transfer relationships and SW coefficients. This is consistent with previous deep-learning-based LST retrieval studies trained on simulation samples (Wang et al., 2021). However, when the validation scenario shifted to real site observations, the performance gap between the two architectures gradually increased. In the site-based leave-one-out cross-validation, PCD-Net achieved an MAE of 1.84 K, lower than the 2.04 K of M-Net. In the independent site validation, the MAE and bias of PCD-Net were 2.45 K and −0.18 K, respectively, also outperforming M-Net, which had an MAE of 2.63 K and a bias of −0.54 K. These results indicate that the explicit component decoupled structure improves model adaptability to real observational conditions and maintains more stable retrieval performance during regional transfer in independent site validation.

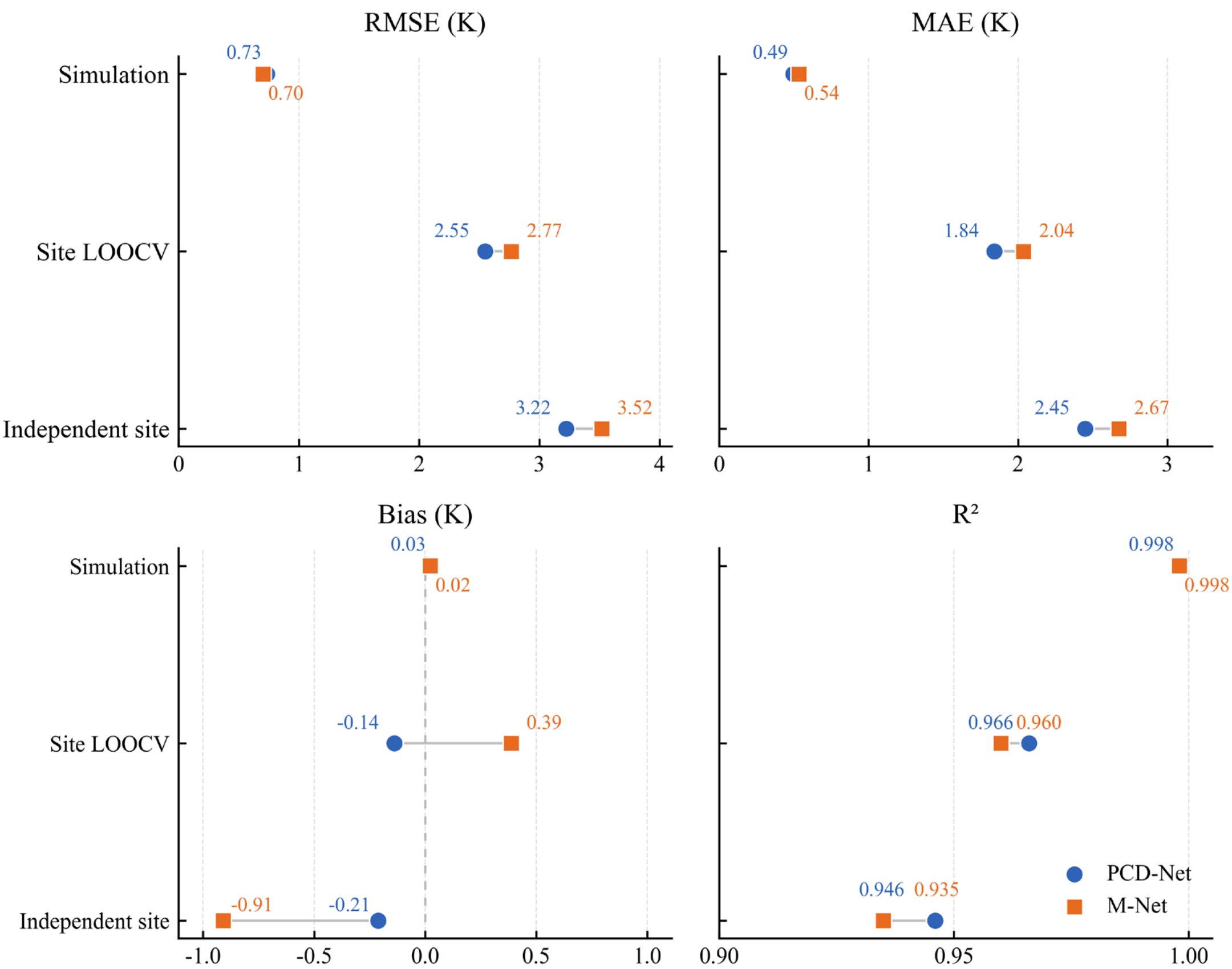

Fig. 15. Performance comparison between PCD-Net and M-Net under different validation strategies. Paired comparisons are shown for RMSE, MAE, bias, and $R^2$. From top to bottom, the rows represent simulation data cross-validation, site-based leave-one-out cross-validation, and independent site validation.

This performance gap may be related to differences in architectural design. In M-Net, multiple SW coefficients are simultaneously output by the same backbone network, resulting in highly shared feature representations. This can implicitly couple coefficients with different physical response characteristics, such as $c_0$ to $c_6$. As a result, uncertainties in input radiometric calibration or noise in auxiliary atmospheric data may propagate through the network in a coupled manner, ultimately affecting LST retrieval accuracy.

In contrast, PCD-Net explicitly decouples the solution pathways of physical coefficients using multiple parallel subnetworks, with each subnetwork responsible for a specific physical component. This design reduces interference among different physical components within a unified feature space. It also helps weaken coupled error propagation across multiple components and improves the interpretability and stability of the retrieval process. This result is consistent with recent studies emphasizing the embedding of physical priors into network structures to enhance model interpretability and robustness (Reichstein et al., 2019).

### 5.3 Limitations and future perspectives

Although PCD-Net achieved good accuracy, stability, and generalization capability in simulation experiments, global site validation, extreme scenario testing, and regional spatial comparisons, its further application is still constrained by common limitations in thermal infrared LST retrieval. First, uncertainties in input variables remain an important source of retrieval errors. The sensitivity analysis showed that errors in at-sensor radiance are the dominant factor causing LST bias. This indicates that high accuracy thermal infrared radiometric calibration, cross-scenario consistency correction, and stable acquisition of auxiliary parameters remain fundamental prerequisites for improving medium- to high-resolution LST retrieval accuracy (Li et al., 2013; Cristóbal et al., 2009). Second, PCD-Net is still limited by the clear-sky observation requirement of thermal infrared remote sensing. Because thermal infrared sensors cannot penetrate clouds, observation gaps are unavoidable in cloudy and persistently cloud covered regions. This affects the spatiotemporal continuity of LST products and limits their application in long term continuous monitoring tasks (Zhang et al., 2020; Duan et al., 2017).

Future research can be advanced in two directions. First, uncertainty propagation analysis, causal inference, and interpretable learning methods can be further introduced to quantitatively characterize the contribution mechanisms of key variables, including radiance, emissivity, water vapor, and path radiance, to LST retrieval errors. This would improve the interpretability of complex coupling processes and enhance the diagnostic capability of retrieval errors (Runge et al., 2019). Second, future work can explore the integration of passive microwave observations, higher-spatiotemporal-resolution reanalysis data, and geostationary satellite thermal infrared observations to construct a multi-source collaborative framework for all-weather and spatiotemporally continuous LST monitoring. The cross-sensor adaptability of PCD-Net among Landsat series sensors, ASTER, ECOSTRESS, and other thermal infrared platforms should also be further strengthened (Quan et al., 2023; Ma et al., 2022; Duan et al., 2017). With continued improvements in structured physical priors, multi-source data fusion, and cross-sensor adaptability, PCD-Net is expected to provide more stable and refined medium- to high-resolution LST data support for global climate change monitoring, agricultural drought assessment, and urban thermal environment studies.

## 6. Conclusion

This study developed and validated PCD-Net, a mechanism–learning coupled framework for land surface temperature retrieval. Using the SW algorithm as the physical backbone, PCD-Net explicitly decomposes LST retrieval into a constant term, a first-order brightness temperature difference term, a second-order brightness temperature difference term, and a coupling residual term. Parallel subnetworks are then used to learn the dynamic coefficients corresponding to different physical components. In terms of training strategy, PCD-Net first uses radiative transfer simulation samples to complete component subnetwork initialization and simulation-domain end-to-end joint optimization, and is then fine-tuned in a supervised manner using global site observation samples. Multi level evaluations based on simulation data and global site observations demonstrate that PCD-Net outperforms traditional mechanistic models and conventional ML baselines in terms of accuracy, stability, and generalization capability, and can adapt to LST retrieval under different climatic backgrounds and diverse underlying surface conditions.

Further analyses show that PCD-Net maintains good stability under complex scenarios, including extremely low temperature, high-temperature, low water vapor, and high water vapor conditions. It alleviates the degradation of traditional mechanism models under boundary conditions and reduces the fluctuations and biases of machine learning methods in sparsely sampled intervals. Meanwhile, the sensitivity analysis indicates that PCD-Net exhibits more stable error responses and lower error dispersion under perturbations in land surface emissivity, atmospheric water vapor, and satellite at-sensor radiance, demonstrating good numerical robustness.

Overall, PCD-Net provides a coupled modeling strategy with both physical interpretability and generalization capability for high accuracy medium- to high-resolution LST retrieval under globally complex environments. Future research can further extend this framework toward all-weather monitoring, multi-source data fusion, and cross-sensor adaptation, thereby enhancing its practical value in global climate change monitoring, agricultural drought warning, and urban thermal environment assessment.

## CRediT authorship contribution statement

**Tian Xie:** Data curation, Investigation, Methodology, Visualization, Writing – original draft, Writing – review & editing. **Menghui Jiang:** Investigation, Methodology, Writing – review & editing. **Chao Zeng:** Methodology, Validation, Writing – review & editing. **Huifang Li:** Project administration, Supervision. **Guanhao Zhang:** Validation, Visualization. **Chan Li:** Data curation, Investigation. **Huanfeng Shen:** Conceptualization, Funding acquisition, Methodology, Project administration, Supervision, Writing – review & editing.

## Declaration of competing interest

The authors declare that they have no known competing financial interests or personal relationships that could have appeared to influence the work reported in this paper.

## Acknowledgements

This study was supported by the Key Program of the National Natural Science Foundation of China under grant 42130108, the National Natural Science Foundation of China under grant 42571435, and the Joint Fund of the National Natural Science Foundation of China under grant U23A2021. The numerical calculations in this paper have been done on the supercomputing system in the Supercomputing Center of Wuhan University.